%% file: manuscript.tex
\documentclass[12pt]{article}

\input{preamble}
\title{\fontsize{18pt}
{18pt}\selectfont Efficient Estimation of Causal Effects Under Two-Phase Sampling with Error-Prone Outcome and Treatment Measurements}

\author{\fontsize{12pt}
{12pt}\selectfont Keith Barnatchez$^1$, Kevin P.~Josey$^2$, Nima S.~Hejazi$^1$, \\ \fontsize{12pt}
{12pt}\selectfont Bryan E.~Shepherd$^3$, Giovanni Parmigiani$^{1,4}$, Rachel C. Nethery$^1$}
\date{\fontsize{12pt}{12pt}\selectfont
$^1$Department of Biostatistics, Harvard T.H.~Chan School of Public Health \\
$^2$Department of Biostatistics and Informatics, Colorado School of Public Health\\
$^3$Department of Biostatistics, Vanderbilt University Medical Center
\\
$^4$Department of Data Science, Dana-Farber Cancer Institute}

\begin{document}
 \maketitle

 \begin{abstract}
Measurement error is a common challenge for causal inference studies using electronic health record (EHR) data,  where clinical outcomes and treatments are frequently mismeasured. Researchers often address measurement error by conducting manual chart reviews to validate measurements in a subset of the full EHR data---a form of two-phase sampling. To improve efficiency, phase-two samples are often collected in a biased manner dependent on the patients' initial, error-prone measurements. In this work, motivated by our aim of performing causal inference with error-prone outcome and treatment measurements under two-phase sampling, we develop solutions applicable to both this specific problem and the broader problem of causal inference with two-phase samples. For our specific measurement error problem, we construct two asymptotically equivalent doubly-robust estimators of the average treatment effect and demonstrate how these estimators arise from two previously disconnected approaches to constructing efficient estimators in general two-phase sampling settings. We document various sources of instability affecting estimators from each approach and propose modifications that can considerably improve finite sample performance in any two-phase sampling context. We demonstrate the utility of our proposed methods through simulation studies and an illustrative example assessing effects of antiretroviral therapy on occurrence of AIDS-defining events in patients with HIV from the Vanderbilt Comprehensive Care Clinic.
\end{abstract}

\thispagestyle{empty}
\setcounter{page}{0}
\clearpage

\newcommand{\showfig}{1}

\setstretch{1.55}

\section{Introduction}
\normalsize
The use of large, administrative electronic health record databases to assess causal relationships has surged over the past few decades. These databases typically contain a vast number of subjects and covariates, offering the potential to address a wide range of scientific questions. However, despite their growing popularity, EHR databases present unique challenges for researchers focused on causal inference questions. Among these challenges is the presence of \textit{measurement error} in key clinical variables, particularly in exposures and outcomes of interest, the two key components of any causal analysis.   
\\ \\
Statistical methods for addressing measurement error have a well-established literature in the parametric modeling tradition \citep{carroll2006measurement} and have gained attention more recently in causal inference (\citealt{valeri2021measurement,barnatchez2024flexible}). A long-established design-based approach to addressing measurement error is to employ a \textit{two-phase sampling} procedure \citep{neyman1938contribution}. In two-phase sampling designs, gold-standard measurements for error-prone variables are obtained for a small subset of the full  data, effectively reframing the measurement error issue as a missing data problem \citep{lotspeich2022efficient}. This small validated subset is often referred to as the \textit{phase-two} sample, while the full dataset is typically termed the phase-one sample.
There is a growing body of work at the intersection of causal inference, measurement error, and missing data that accommodates sampling schemes of this nature \citep[e.g.,][]{kallus2024role,kennedy2020efficient,levis2024double}. Most closely related to our work, recent developments from the semi-supervised learning literature explore variations of the measurement error problem, when the phase-two validation data is obtained completely at random but at a rate that decays to zero as the size of the overall dataset grows \citep{hou2025efficient}. 
\\ \\
While these methods serve as important advances in the careful use of EHR data for drawing causal inferences, several practical challenges remain unaddressed. We highlight two key challenges in this study. First, specific to our applied science motivation, there is a need for semi-parametric efficient methods in measurement error settings that explicitly account for \textit{biased} phase-two sampling schemes. Biased sampling occurs when phase-two sampling probabilities depend on measured covariates \textit{and} the error-prone measurements of the outcome and treatments, potentially resulting in a validation sample that is not representative of the target population from which the EHR data is drawn. Although such sampling requires careful statistical methodology, biased sampling rules can offer significant efficiency gains for downstream estimation tasks compared to simple random validation sampling \citep{breslow1999design}, especially in scenarios where the outcome and exposure of interest are rare. In practice, budgetary constraints often necessitate that only a small share of EHR data can be validated, meaning biased sampling rules can, in terms of efficiency of the resulting estimator, drastically increase the effective sample size of the phase-two data. The use of biased sampling schemes to address measurement error is well-studied in the context of parametric and semi-parametric models that aim to capture associations between variables \citep{shepherd2023multiwave}, but is relatively underexplored in settings where interest lies in semi-parametric efficient estimation of causal effects defined non-parametrically.
\\ \\
The second challenge applies more generally to the problem of performing semi-parametric efficient estimation under two-phase sampling schemes, of which our specific problem of interest is a special case. We document that there are two general approaches to performing causal inference under two-phase sampling schemes. In the first approach, the researcher uses a combination of unconfoundedness and missing-at-random assumptions to identify the causal quantity of interest as a statistical functional of the observed data distribution. They then use standard tools from semi-parametric theory to derive efficient estimators of this functional. This is undoubtedly the approach most commonly taken to derive asymptotically efficient estimators in causal inference applications and has been taken to address several problems arising at the intersection of causal inference and missing data \citep{kennedy2020efficient,levis2024robust} Alternatively, the second approach leverages links between the \textit{observed} data distribution, and the underlying \textit{complete} data distribution one would have access to if it were possible to observe gold-standard measurements for every subject \citep{rose2011targeted,hejazi2021efficient,wang2023maximin}. We document that the construction of semi-parametric efficient estimators tends to be drastically simplified under the second approach. While the estimators arising from both approaches will be asymptotically equivalent in general two-phase sampling problems, there is no consensus on how their behavior differs in finite samples. This issue is further compounded by the fact that in practice, researchers tend to implement one of the two approaches without carefully considering the merits of the other. Further understanding of the finite-sample behavior of estimators arising from each approach is crucial, since  validated sample sizes will tend to be small in practice. 
\\ \\
In this paper, we address these two challenges by deriving 
semi-parametric efficient estimators of counterfactual mean outcomes under settings where (1) the outcome and exposure are measured with error, and (2) the researcher is able to collect gold-standard measurements for the outcome and exposure for a small but \textit{biased} subsample of the overall dataset. We present two classes of estimators, one arising from each of the two general approaches to performing semi-parametric causal inference under two-phase sampling described above. 
Through simulations and a study based on data from the Vanderbilt Comprehensive Care Clinic (VCCC) we show that while these estimators are asymptotically equivalent, their behavior can yield meaningfully different results in finite samples and in settings where the phase-two sample is relatively small. 
To address challenges that typically arise with small amounts of phase-two data, we present modifications based on empirical efficiency maximization \citep{rubin2008empirical} that can dramatically improve efficiency of the individual estimators in finite samples. Further, we present an ensemble estimator which optimally combines these two estimators to maximize finite-sample efficiency, while retaining the asymptotic distribution shared by the two estimators.
\\ \\
The remainder of this paper is structured as follows.  
Section~\ref{prob-setting} frames the problem setting of interest, as well as accompanying assumptions and causal estimands, while in Section~\ref{sec:id-eff-theory} we present identification results and efficiency theory under the two sets of general approaches one can take in two-phase designs. In Section \ref{sec:methods}, we introduce corresponding one-step estimators
and detail their asymptotic properties. Additionally, we present minor modifications to our estimation strategies which improve their finite sample performance.
We explore the finite-sample characteristics of our proposed methods in Section~\ref{simmy}, turning our attention to their performance on data from the VCCC in Section~\ref{data-application}. Finally, in Section~\ref{discussion} we conclude with a discussion of our findings and directions for future research.

\section{Problem Setting}\label{prob-setting}

\subsection{Data Structure}
\label{data-structure}

Following \cite{kallus2024role} and \cite{hou2025efficient}, we frame our problem through a missing data framework. Suppose we observe $n$ independent samples
\begin{equation}
    \bO_i = (R_i Y_i, Y_i^*, R_i A_i, A_i^*, \bX_i, R_i) \overset{\text{iid}}{\sim} \Pd, \ i \in \{1,\dots,n\} \ 
\end{equation}
where $Y_i$ and $A_i$ are an outcome and exposure of interest accompanied by the error-prone measurements $Y_i^*$ and $A_i^*$, and $\bX_i$ is a vector of subject-level covariates. Critical to our setting, $A_i$ and $Y_i$ are only observed when the phase-two validation indicator $R_i = 1$ and are missing otherwise. For compactness, throughout the manuscript we let $\bW \eqdef (A^*, Y^*)$ collect the error-prone measurements and $\bZ \eqdef (\bX, \bW)$ denote the variables that are always observed, regardless of membership in the phase-two sample. One can additionally conceptualize observations arising from the \textit{complete-data} distribution $(Y_i, Y_i^*, A_i, A_i^*, \bX_i, R_i) \sim\Pd^\text{C}$, where $Y_i$ and $A_i$ are available for all subjects. The observed data distribution $\Pd$ can be viewed as a coarsening of $\Pd^\text{C}$, the latter of which we cannot directly sample from. The approaches we consider will make reference to both $\Pd$ and $\Pd^\text{C}$, and we occasionally subscript expectations $\E$ to indicate which distribution the expectation is taken over.
\\ \\
We let $Y_i(1)$ and $Y_i(0)$ denote subject $i$'s potential outcomes under a treated and control exposure, respectively. Our ultimate interest lies in estimating the average treatment effect (ATE) $\psi = \E[Y(1) - Y(0)]$, where we define $\psi_a = \E[Y(a)]$ so that $\psi=\psi_1-\psi_0$.
Estimation of $\psi$ in our setting requires the estimation of numerous nuisance functions, which we define in Table \ref{tab:nuis-funcs}.

\ifthenelse{\equal{\showfig}{1}}{
\begin{table}[h!]
    \centering
    \begin{tabular}{lll}
    \toprule
     Nuisance function    & Definition & Interpretation  \\
     \midrule
     $\mu_a(z)$    & $\E_\Pd[Y|\bZ = z, R=1]$ & Outcome imputation model \\
     $\lambda_a(z)$ & $\PP_\Pd[A=a|\bZ=z,R=1]$ & Treatment imputation model \\
     $\eta_a(x)$ & $\E_\Pd[\lambda_a(\bZ) \cdot \mu_a(\bZ) | \bX=x]$ & Marginalized outcome imputation \\
     $\pi_a(x)$ & $\E_\Pd[\lambda_a(\bZ) | \bX=x]$ & Imputed propensity score \\
     $\kappa(z)$ & $\PP_\Pd(R=1|\bZ=z)$ & Phase-two selection  model \\
     $m_a(x)$ & $\E_{\Pd^\text{C}}[Y|A=a,\bX=x]$ & Full data outcome regression \\
     $g_a(x)$ & $\PP_{\Pd^\text{C}}(A=a|\bX=x)$ & Full data propensity score \\
    \bottomrule
    \end{tabular}
    \caption{Nuisance function definitions.}
    \label{tab:nuis-funcs}
\end{table}}

\vspace{-1.5em}
\subsection{Assumptions}
\label{identification}
In this section, we discuss conditions that allow for the identification of $\psi$ from the observed data. We begin by invoking a set of core assumptions commonly made in observational causal inference.
\begin{assumption}[SUTVA]
\label{sutva1}
$Y = AY(1) + (1-A)Y(0)$. Further, $(Y_i(1), Y_i(0)) \indep A_j \ \ \forall i \neq j$
\end{assumption}
\begin{assumption}[Positivity]
\label{positivity2}
 $0<\mathbb{P}_\Pd(A=1|\bX=x)<1$ for all $x$ with positive support
\end{assumption}
\begin{assumption}[Unconfoundedness]
\label{unconfoundedness3}
 $Y(a) \indep A | \bX$ 
\end{assumption}
Assumptions \ref{sutva1} through \ref{unconfoundedness3} are sufficient for identifying the average treatment effect in the population corresponding to the phase-two validation data. To account for the possibility that the availability of phase-two data systematically depends on the confounders $\bX$ and error-prone exposure and outcome measurements, we make two additional assumptions:
\begin{assumption}[Outcome and Exposure Missing  At Random]
\label{ymcar4}
$(Y,A)  \indep R | \bZ$
\end{assumption}
\begin{assumption}[Positivity of validation data selection]
\label{val-positivity6}
$0 < \kappa(z) <1$ for all $z$ with positive support
\end{assumption}
Figure \ref{swig} illustrates a causal diagram consistent with independence Assumptions \ref{unconfoundedness3} and \ref{ymcar4}.
\ifthenelse{\equal{\showfig}{1}}{
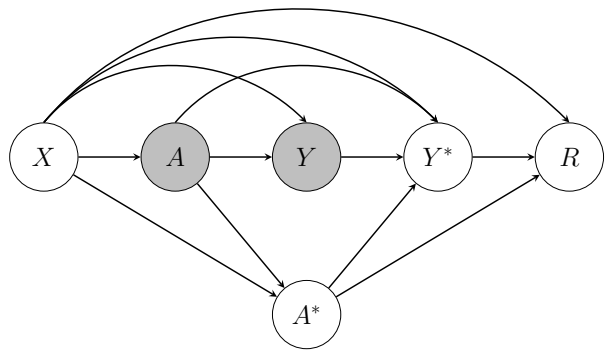
\begin{figure}[h!]
    \centering
    \resizebox{0.35\textheight}{!}{
    \fontsize{14}{14}\selectfont
    \begin{tikzpicture}[node distance=0.8cm]
    \node[circle,draw,minimum size=1.3cm] (X) at (-3,0) {$X$};
    \node[circle,draw,minimum size=1.3cm,fill=lightgray] (A) at (-0.5,0){$A$};
    \node[circle,draw,minimum size=1.3cm,fill=lightgray] (Y) at (2,0) {$Y$};
    \node[circle,draw,minimum size=1.3cm] (Astar) at (2,-3) {$A^*$};
    \node[circle,draw,minimum size=1.3cm] (Ystar) at (4.5,0) {$Y^*$};
    \node[circle,draw,minimum size=1.3cm] (S) at (7,0) {$R$};

    \draw[->, thick, >=stealth] (A) -- (Astar);
    \draw[->, thick, >=stealth,opacity=1] (X.north) to [out=50] (Ystar.north);
    \draw[->, thick, >=stealth,opacity=1] (A.north) to [out=50] (Ystar.north);
    \draw[->, thick, >=stealth] (X) -- (A);
    \draw[->, thick, >=stealth] (X) -- (Astar);
    \draw[->, thick, >=stealth] (A) -- (Y);
    \draw[->, thick, >=stealth] (X.north) to [out=50] (Y.north);
    \draw [->, thick, >=stealth,opacity=1] (X.north) to [out=50] (S.north);
    \draw[->, thick, >=stealth] (Y) -- (Ystar);
    \draw[->, thick, >=stealth] (Astar) -- (S);
    \draw[->, thick, >=stealth] (Astar) -- (Ystar);
     \draw[->, thick, >=stealth] (Ystar) -- (S);
\end{tikzpicture}}
\caption{A causal diagram consistent with independence Assumptions \ref{unconfoundedness3} and \ref{ymcar4}. The shaded random variables---corresponding to the exposure and outcome of interest---are only observed when the validation indicator $R=1$. Selection into the phase-two data is determined \textit{after} observing $\bX$, $A^*$ and $Y^*$.}
\label{swig}
\end{figure}}

Assumption \ref{ymcar4} is analogous to missing at random (MAR) assumptions that frequent the missing data literature, while Assumption \ref{val-positivity6} ensures overlap between covariate distributions in the validation sample and the broader target population.  Further, in the Supplementary Materials we show that $\Pd^\text{C}$ is identified by $\Pd$ under Assumptions \ref{ymcar4} and \ref{val-positivity6}, implying one can identify any quantity dependent on $\Pd^\text{C}$ through the observed data. We make use of this result in Section \ref{sec:approach2}. 
\\ \\
In two-phase sampling studies, Assumptions \ref{ymcar4} and \ref{val-positivity6} can be enforced by design as the researcher will often have control over the phase-two sampling mechanism. While the complete-case probabilities $\kappa(\bm Z)$ are often known in two-phase studies, we explore methods that are agnostic to whether  $\kappa(\bm Z)$ is known or needs to be estimated in order to account for broader sampling regimes. 

\section{Identification and Efficiency Theory}
\label{sec:id-eff-theory}

In related papers at the intersection of causal inference and missing data, researchers have taken two general approaches to obtain identifying statistical functionals of causal quantities, and the corresponding efficient influence curves (EICs) of these functionals. The first general approach operates on the observed data structure and is the standard approach taken in causal inference problems. A second general approach, which has recently enjoyed increased attention in causal inference studies (\citealt{kennedy2016semiparametric,hejazi2021efficient,hou2025efficient,wang2023maximin}), leverages links between the observed data distribution and the underlying complete data distribution that one would have access to in the absence of any missing data. While we will show that the estimators arising from these approaches are asymptotically equivalent, their behavior can meaningfully differ in finite samples. Before examining their relative merits in Sections \ref{simmy} and \ref{data-application}, we outline both approaches in this section. 

\subsection{Approach 1: Using the Observed Data Structure}
\label{sec:approach1}

Given the observed data distribution $\Pd$, one can make use of Assumptions \ref{sutva1}-\ref{val-positivity6} to obtain non-parametrically identifiable functionals of $\E[Y(a)]$. We emphasize that this approach to identification is standard in causal inference. The following Theorem provides an identifying expression for $\psi_a$ derived from this approach.
\begin{theorem}
\label{id-thm}
    Under Assumptions \ref{sutva1}-\ref{val-positivity6}, for $a \in \{0,1\}$,
    \begin{equation}
        \label{id-exp}
        \psi_a =
         \E_\Pd\left[\frac{\eta_a(\bX)}{\pi_a(\bX)}\right] = \E_\Pd \left[\frac{\E_\Pd(\lambda_a(\bZ) \cdot \mu_a(\bZ) | \bX)}{\E_\Pd(\lambda_a(\bZ) | \bX)} \right].
    \end{equation}
\end{theorem}
The corresponding proof of Theorem \ref{id-thm}, and all theorems that follow, can be found in the Supplementary Materials.
The identifying expression \eqref{id-exp} is similar to those found in \cite{kennedy2020efficient} and \cite{kallus2024role}, and can be viewed as an inverse probability weighted estimator of the imputed values of $Y$ and $A$ after marginalizing out the error-prone measurements $\bW$ used to form the imputations. Critically, Theorem \ref{id-thm} implies one can obtain a plug-in estimate for $\psi_a$ by: (1) obtaining nuisance model estimates $\hat \lambda_a(\bZ)$ and $\hat\mu_a(\bZ)$ via estimation using the validated phase-two data;  
(2) regressing $\hat \lambda_a(\bZ)\cdot\hat\mu_a(\bZ)$ onto $\bX$, as well as $\hat \lambda_a(\bZ)$ onto $\bX$ in the full dataset to obtain the nuisance model estimates $\hat \eta_a(\bX)$ and  $\hat \pi_a(\bX)$, and (3)
constructing the final plug-in estimate for $\psi_a$ as
\begin{equation}
\label{plug-in-eq}
\hat \psi_a^\text{PI,1} \eqdef \frac{1}{n} \sum_{i=1}^n \frac{\hat \eta_a(\bX_i)}{\hat \pi_a(\bX_i)} \ .
\end{equation}
While the plug-in estimator \eqref{plug-in-eq} is consistent under correct model specification, it is well known that it will only achieve desired parametric rates of convergence if one correctly specifies statistically consistent models for the estimation of all nuisance functions (\citealt{kennedy2024semiparametric}), making statistical inference an intractable task.
In general semi-parametric estimation problems, when one wishes to avoid making possibly unjustified parametric modeling assumptions, one can mitigate this plug-in bias by incorporating estimators based on the efficient influence curve (EIC) for the corresponding target functional \citep{robins1994estimation, tsiatis2006semiparametric}. The EIC for $\psi_a$, a crucial ingredient for constructing efficient non-parametric estimators which can be derived by using standard tools from semi-parametric theory, is provided below.
\begin{theorem}[Semi-parametric Efficiency Bound]
\label{if-thm}
The efficient influence curve, $\phi_a(\bm O,\Pd)$, of $\psi_a$ from
Theorem \ref{id-thm} is
\begin{align}
\phi_a(\bm O,\Pd) =&~
\frac{\eta_a(\bX)}{\pi_a(\bX)} - \psi_a + \frac{\lambda_a(\bZ)}{\pi_a(\bX)} \left(
\mu_a(\bZ)
- \frac{\eta_a(\bX)}{\pi_a(\bX)}
 \right) \nonumber
\\
&+
 \frac{R I(A=a)}{\kappa(\bZ)\pi_a(\bX)}\left(Y-\frac{\eta_a(\bX)}{\pi_a(\bX)}\right) 
- \frac{R\lambda_a(\bZ) }{\kappa(\bZ)\pi_a(\bX)}\left(
\mu_a(\bZ)
- \frac{\eta_a(\bX)}{\pi_a(\bX)}
 \right) \ . \label{eq:eif-thm-eq-approach1}
\end{align}
In turn, the semi-parametric efficiency bound for estimating $\psi_a$ is given by $\E_\Pd(\phi_a(\bm O,\Pd)^2)$.
\end{theorem}
Notably, the efficiency bound $\E[\phi_a(\bm O, \Pd)]^2$ depends on not only $\eta_a(\bX)$ and $\pi_a(\bX)$, but also the imputation models $\lambda_a(\bZ)$ and $\mu_a(\bZ)$ and sampling probabilities $\kappa(\bZ)$. 
Along with characterizing the best-case asymptotic variance for any regular asymptotically linear (RAL) estimator of $\psi_a$ in the non-parametric model $\mathcal{M}$, Theorem \ref{if-thm} provides a means for constructing efficient semi-parametric estimators of $\psi_a$ under the missing data structure presented in Section \ref{data-structure}, provided that Assumptions \ref{sutva1}-\ref{val-positivity6} hold. The construction of such estimators is detailed in Section \ref{sec:methods}.

\subsection{Approach 2: Leveraging the Complete Data Distribution}
\label{sec:approach2}

The second general approach, which we refer to  as Approach 2, was initially suggested by \cite{robins1994estimation} and extensively developed in \cite{laan2003unified} and is increasingly used for two-phase sampling problems \citep{rose2011targeted, hejazi2021efficient}. 
Approach 2 begins by considering the analysis one would conduct in the \textit{absence of missingness} of $Y$ and $A$. To be concrete, recall the \textit{complete data} distribution that an analyst would ideally possess $\Pd^\text{C}$,
where $Y$ and $A$ are observed for \textit{all} subjects. The  observed data distribution $\Pd$ can be viewed as a coarsening of the complete data distribution $\Pd^\text{C}$.
\\ \\
In the absence of missing data, it is well-established that one can identify $\E[Y(a)]$ by the g-formula functional (\citealt{hernan2024causal})
\begin{equation}
\label{eq:complete-id}
\psi_a = \E_{\Pd^\text{C}}[m_a(\bX)],
\end{equation}
where $m_a(\bX)\eqdef \E_{\Pd^\text{C}}[Y|A=a,\bX]$ is the complete data conditional outcome regression for treatment level $A=a$. 
Notice $m_a(\bX)$ can be consistently estimated through a regression of $Y$ on $\bX$ among subjects with $A=a$ that weights observations by $R/\hat \kappa(\bZ)$. \cite{rose2011targeted} have established that with appropriate function classes, such an estimation strategy is consistent under the exchangeability and MAR assumptions, and can similarly be used to estimate the full-data propensity scores  and $g_a(\bX) \eqdef \PP(A=a|\bX)$. This estimation strategy suggests the alternative plug-in estimator
\begin{equation}
\label{eq:alt-plugin}
    \hat \psi_a^\text{PI,2} = \frac{1}{n}\sum_{i=1}^n \hat m_a(\bX_i),
\end{equation}
where $\hat m_a(\bX)$ is obtained through a  regression that assigns weights $R/\hat \kappa(\bZ)$ to each subject.  
Though appealing due to its parsimony, $\hat \psi_a^\text{PI,2}$ is subject to the same drawbacks suffered by $\hat \psi_a^\text{PI,1}$, necessitating corrections based on the efficient influence curve under the observed data structure.
While the efficient influence curve for (\ref{eq:complete-id}) under the complete data distribution is given by
\[
\chi_a(\bm O;\Pd^\text{C}) = \frac{I(A=a)}{g_a(\bX)}(Y-m_A(\bX)) + m_a(\bX) - \psi_a \ ,
\]
in practice we require the efficient influence curve under the \textit{observed} data distribution. 
A crucial result for \textit{general} two-phase sampling settings \citep{rose2011targeted,hejazi2021efficient,levis2024double} links the representation from Theorem \ref{if-thm} to an alternative representation written in terms of $\chi_a(\bm O;\Pd^\text{C})$. Specifically, 
letting $\varphi_a(\bZ) \eqdef \E_\Pd[\chi_a(\bm O;\Pd^\text{C}) | \bm Z, R=1]$, we have the following result:
\begin{proposition}
\label{prop:eif-prop}
Under Assumptions \ref{sutva1}-\ref{val-positivity6}, the Approach 1 efficient influence curve $\phi_a(\bm O,\Pd)$ can equivalently be written as
\begin{equation}
\label{eq:alt-eif}
\phi_a(\bm O;\Pd) = \phi_a^\text{ALT}(\bm O;\Pd) \eqdef \frac{R\chi_a(\bm O;\Pd^\text{C})}{\kappa(\bm Z)} - \left( \frac{R}{\kappa(\bm Z)} - 1\right) \varphi_a(\bZ).
\end{equation}
\end{proposition}
Critically, (\ref{eq:alt-eif}) implies that along with the representation provided in Theorem \ref{if-thm}, the EIC for $\psi_a$ under the observed data distribution $\Pd$ can equivalently be written as a function of the \textit{complete}-data EIC and the validation sampling probabilities $\kappa(\bZ)$. Notably, the influence curve in Proposition \ref{prop:eif-prop} closely resembles the form of the complete-data curve \( \chi_a(\bm O; \Pd^C) \), and can be viewed as the efficient influence curve for a missing data functional in which the usual outcome $Y$ is replaced by the pseudo-outcome $\chi_a(\bm O, \Pd^\text{C}$).
While estimators constructed from either $\phi_a(\bm O;\Pd)$ or $\phi_a^\text{ALT}(\bm O;\Pd)$ will be asymptotically equivalent under standard regularity conditions outlined in the Supplementary Materials, their finite sample behavior may meaningfully differ due to numerous factors. We highlight these factors in the next section. 

\section{Efficient Estimation}
\label{sec:methods}

Given the two distinct plug-in estimators suggested by these two identification strategies---and corresponding efficient influence curve representations---one can use standard tools from semi-parametric theory to correct the first-order bias of the plug-in estimators. In particular, we consider one-step bias-corrected estimators, whose construction involves adding the empirical average of the estimated efficient influence curve on to the initial plug-in estimator \citep{pfanzagl1985contributions}. We primarily focus on one-step estimators as their implementations enable modifications that can significantly improve their finite-sample performance in two-phase sampling settings. We discuss one such modification in Section~\ref{extensions}. In the Supplementary Materials we outline an efficient estimator of $\psi_a$ based on targeted maximum likelihood estimation (TMLE), an alternative framework which produces estimators asymptotically equivalent to those produced by the one-step estimation framework \citep{kennedy2024semiparametric}. 
\\ \\
Given the one-step bias correction strategy, the proposed estimators take the  form
\newcommand{\sumn}{\frac{1}{n}\sum_{i=1}^n}
\begin{align}
    \hat \psi_a^\text{OS,1} &\eqdef \hat \psi_a^\text{PI,1} + \sumn
    \bigg\{
 \frac{\hat \eta_a(\bX_i)}{\hat\pi_a(\bX_i)} + \frac{\hat\lambda_a(\bZ_i)}{\hat\pi_a(\bX_i)} \left(\hat \mu_a(\bZ_i) - \frac{\hat \eta_a(\bX_i)}{\hat\pi_a(\bX_i)} \right) + 
 \frac{R_i I(A_i=a)}{\hat\kappa(\bZ_i)\hat\pi_a(\bX_i)}\left(Y-\frac{\hat\eta_a(\bX_i)}{\hat\pi_a(\bX_i)}\right)
\nonumber \\
& \ \ \ \ \ \ \ \ \ \ \ \  - 
\frac{R_i\hat\lambda_a(\bZ_i) }{\hat\kappa(\bZ_i)\hat\pi_a(\bX_i)}\left(
\hat\mu_a(\bZ_i)
- \frac{\hat\eta_a(\bX_i)}{\hat\pi_a(\bX_i)}
 \right)  - \hat \psi_a^\text{PI,1} \bigg\}; \label{eq:dr-os} 
 \\
 \hat \psi_a^{\text{OS,2}} &= \hat \psi_a^\text{PI,2} + \frac{1}{n}\sum_{i=1}^n \bigg \{ \frac{R_i}{\hat \kappa(\bZ_i)}\left(\frac{I(A_i=a)}{\hat g_a(\bX_i)}(Y-\hat m_{A_i}(\bX_i)) + \hat m_a(\bX_i)  - \hat \psi_a^\text{PI,2}\right) \nonumber \\
& \ \ \ \  \ \ \  \ \ \ \  \ \ \ \ \ \ \ \  \ \ -  \left( \frac{R_i}{\hat \kappa(\bm Z_i)} - 1\right) \hat \varphi_a(\bZ_i) \bigg \}. \label{eq:dr-os-alt}
\end{align}
In settings where the phase-two sampling probabilities are known, one can simply assign $\hat\kappa(\bm Z) = \kappa(\bZ)$, though we recommend estimation to allow for further efficiency gains \citep{tsiatis2006semiparametric, rose2011targeted}. Estimation can be performed in these settings by fitting a logistic regression of $R$ on $\bm Z$ while including an offset term for the true log-odds of sampling. 
In settings where $\kappa(\bZ)$ is unknown, we recommend the use of flexible methods for its estimation. For such settings, we provide recommendations on best practices for estimating $\varphi_a(\bm Z)$ in Section \ref{extensions}.

\subsection{Theoretical Results}
While $\hat \psi_a^\text{OS,1}$ and $\hat \psi_a^\text{OS,2}$ will be asymptotically equivalent under correct and sufficiently fast convergence rates of all nuisance models, it is of practical interest to know under which specific conditions each estimator attains consistency and asymptotic normality. The following two theorems outline the conditions required for each estimator.
\begin{theorem}[Asymptotic distribution of the Approach 1 one-step estimator]
\label{dr-theorem}
Suppose $||\hat \phi_a - \phi_a|| = o_\Pd(1)$ and that all nuisance function estimates are obtained from a separate, held-out sample.  Then,
\begin{align*}
\hat \psi_a^\text{OS,1} - \psi_a &= \sumn\phi_a(\bm O, \Pd) +  o_\Pd\left( 
\frac{1}{\sqrt n}\right)  \nonumber \\ &+ O_\Pd \left(  (|| \hat \eta_a - \eta_a ||  + || \hat \pi_a - \pi_a || ) \cdot  || \hat \pi_a - \pi_a ||  +  (|| \hat \lambda_a - \lambda_a  || + ||\hat \mu_a - \mu_a ||) \cdot || \hat \kappa - \kappa || \right).
\end{align*}
Further, if 
\begin{enumerate}
    \item $|| \hat \eta_a - \eta_a ||\cdot || \hat \pi_a - \pi_a || = o_\Pd(1/\sqrt n)$
    \item $|| \hat \pi_a - \pi_a || = o_\Pd(n^{-1/4})$
    \item $(|| \hat \lambda_a - \lambda_a  || + ||\hat \mu_a - \mu_a ||) \cdot || \hat \kappa - \kappa ||= o_\Pd(1/\sqrt n)$,  
\end{enumerate}
then $\hat \psi_a^\text{OS,1}$ is $\sqrt n$-consistent and asymptotically normal. Additionally, the asymptotic variance of $\hat \psi_a^\text{OS,1}$ equals the semi-parametric efficiency bound $\E[\phi_a(O,\Pd)^2]$.
\end{theorem}
There are multiple immediate consequences of Theorem~\ref{dr-theorem}. First, notice that sufficiently fast $n^{-1/4}$ consistent estimation of the imputed propensity score $\pi_a$ is \textit{required} for $\sqrt n$-consistency of $\hat \psi_a^\text{OS,1}$, a finding in line with that of \cite{kennedy2020efficient}, who noted a similar consistency requirement in settings with missing treatment information. Second, the above decomposition implies that in settings where the sampling probabilities are known, one does not require consistent estimation of the imputation models $\lambda_a$ or $\mu_a$. However, we note that consistent estimation of $\eta_a$ and $\pi_a$ is unlikely in the event that $\lambda_a$ and $\mu_a$ are inconsistently estimated, suggesting $\hat \psi_a^\text{OS,1}$ requires a relatively strong set of conditions for consistency and  asymptotic normality.
\begin{theorem}[Asymptotic distribution of the Approach 2 one-step estimator]
\label{thm:dr-theorem-alt}
Suppose $\lVert \hat \phi_a^\text{ALT} - \phi_a^\text{ALT} \rVert = o_\Pd(1)$ and that all nuisance function estimates are obtained from a separate, held-out sample. Then
\begin{align*}
\hat \psi_a^\text{OS,2} - \psi_a &= \sumn \phi_a^\text{ALT}(\bm O_i,\Pd) + o_\Pd\left( 
\frac{1}{\sqrt n}\right) \\ &+ O_\Pd \left(  || \hat m_a - m_a || \cdot || \hat g_a - g_a ||  + \ || \hat \varphi_a - \varphi_a  || \cdot || \hat \kappa - \kappa || \right).   
\end{align*}
Further, if 
\begin{enumerate}
    \item $|| \hat m_a - m_a ||\cdot || \hat g_a - g_a || = o_\Pd(1/\sqrt n)$
    \item $|| \hat \varphi_a - \varphi_a  || \cdot || \hat \kappa - \kappa ||= o_\Pd(1/\sqrt n)$,  
\end{enumerate}
then $\hat \psi_a^\text{OS,2}$  is $\sqrt n$-consistent and asymptotically normal.  Additionally, the asymptotic variance of $\hat \psi_a^\text{OS,2}$ attains  the semi-parametric efficiency bound $\E[\phi_a(O,\Pd)^2] = \E[\phi_a^\text{ALT}(O,\Pd)^2]$. 
\end{theorem}
Recall that under correct specification and sufficiently fast estimation of all nuisance functions, $\hat \psi_a^\text{OS,1}$ and $\hat \psi_a^\text{OS,2}$ are asymptotically equivalent, with both estimators achieving the semi-parametric efficiency bound for estimating $\psi_a$. Theorem~\ref{thm:dr-theorem-alt} suggests that relative to $\hat \psi_a^\text{OS,1}$,  $\hat \psi_a^\text{OS,2}$ requires less stringent conditions to attain consistency and asymptotic normality. Particularly in study designs where $\kappa$ is known, we simply require $|| \hat m_a - m_a ||\cdot || \hat g_a - g_a || = o_\Pd(1/\sqrt n)$, which is the usual double-robustness condition that arises in simple cross-sectional observational studies not subject to the complications of two-phase sampling. This result makes explicit a crucial benefit of controlling the phase-two sampling probabilities under Approach 2:  in terms of consistency, estimation of $\psi_a$ in two-phase settings is  no more difficult a task than in settings where one has complete data.

\subsection{Connections Between the Two Approaches}

The efficient influence curve representation in~\eqref{eq:alt-eif} makes explicit the links between the doubly-robust estimators constructed from  Approach 1 and Approach 2. In Approach 2, the full-data conditional ATE and propensity score functions are replaced by the nuisance functions $\eta_a(\bX)/\pi_a(\bX)$ and $\pi_a(\bX)$, respectively. Further, the remaining components of the efficient influence curve in Approach 1 are absorbed into $\varphi_a(\bZ)$, which regresses the full-data efficient influence curve on the variables influencing selection into the phase-two sample. In the Supplementary Materials we show that
\begin{equation}
\label{eq:varphi_a-expansion}
   \varphi_a(\bZ) = \frac{\lambda_a(\bZ) \mu_a(\bZ)}{\pi_a(\bX)}
- \frac{\lambda_a(\bX)\eta_a(\bX)}{\pi_a(\bX)^2} + 
\frac{\eta_a(\bX)}{\pi_a(\bX)} - \psi_a \ .
\end{equation}
In this sense, estimators constructed from Approach 2 possess two major departures from the Approach 1 estimators. First, the Approach 2 estimators target the full-data conditional ATE and propensity score functions through weighted regressions, rather than aiming to reconstruct them through the nuisance functions $\eta_a(\bX)$ and $\pi_a(\bX)$.  Second, and most notably, the Approach 2 estimator effectively collapses the complex composite of nuisance functions (\ref{eq:varphi_a-expansion}) into a single nuisance function $\varphi_a(\bZ)$, reducing the total number of nuisance functions that need to be estimated. 
While these two departures result in a seemingly more attainable set of consistency requirements for the Approach 2 estimators, this comes at the cost of introducing the possibly unwieldy nuisance function $\varphi_a(\bZ)$. \eqref{eq:varphi_a-expansion} demonstrates that $\varphi_a(\bZ)$ is a complex function of \textit{four} underlying nuisance components, suggesting it will be a challenging function to estimate given the typically small sample sizes of phase-two data. Though Theorem~\ref{thm:dr-theorem-alt} demonstrates incorrect estimation of $\varphi_a(\bZ)$ will not result in bias when the two-phase sampling probabilities are known, severe misspecification can  hamper \textit{efficiency} of the corresponding estimator.
\\ \\
While the difficulty of estimating $\varphi_a(\bZ)$ suggests that Approach 2 estimators will tend to be inefficient in finite samples, estimators derived from Approach 1 face similar challenges in finite samples, albeit for different reasons. From (\ref{eq:dr-os}), the construction of the Approach 1 one-step estimator involves numerous second-and third degree products of weight terms, which have been demonstrated to generate instability in generalizability and transportability settings (\citealt{chattopadhyay2024one}). In turn, although Approach 1 avoids direct estimation of $\varphi_a(\bZ)$ through its decomposition in (\ref{eq:varphi_a-expansion}), the resulting decomposition introduces numerous multiplicative weighting terms. These weighting terms will typically introduce instability, effectively reducing the efficiency of the estimator in finite samples.
\\ \\
These findings suggest that while the Approach 1 and Approach 2 estimators are asymptotically equivalent and semi-parametric efficient, both will likely be inefficient in finite samples. Further, the two sources of inefficiency---multiplicative weighting terms for Approach 1, and the inherently complex function $\varphi_a(\bZ)$ in Approach 2---are difficult to compare. Depending on the specific application, Approach 1 estimators may perform better than Approach 2 estimators and vice versa. In the remainder of this section, we demonstrate  modes by which one can leverage the bias structure of the Approach 2 estimators to avoid the drawbacks of outright estimation of $\varphi_a(\bZ)$. Further, we present an ensemble estimator that circumvents the need to choose between the two estimation strategies.

\subsection{Empirical Efficiency Maximization}
\label{extensions}

A key result of Theorem \ref{thm:dr-theorem-alt} implies that when the sampling probabilities $\kappa(\bZ)$ are known or can be estimated well, the asymptotic  bias  of  $\hat \psi_a^\text{OS,2}$ will be solely dictated by the accuracy of $\hat m_a(\bX)$ and $\hat g_a(\bX)$, and unaffected by $\hat \varphi_a(\bZ)$. However, inaccurate estimation of $\varphi_a(\bZ)$ \textit{will} impact the efficiency of the Approach 2 estimators. Given that $\varphi_a(\bZ)$ is often a complicated function in many applications, and the typically small amount of validated data available in practice, Approach 2 estimators may suffer from inflated variance in practical applications.
\\ \\
To address this shortcoming, one can leverage the bias structure presented in Theorem 4 by incorporating ideas from the \textit{empirical efficiency maximization} (EEM) framework (\citealt{rubin2008empirical}). EEM is based upon the observation that under correct specification of $m_a$ and $g_a$, the true $\varphi_a$ will minimize the asymptotic variance of $\hat \psi_a^\text{OS,2}$. Letting $\chi_a(\bm O_i ; \hat P^\text{C}) := \chi_a(\bm O_i ; \hat m_a, \hat g_a)$, one can explicitly target this variance-minimizing property by choosing $\hat \varphi_a$ to minimize the empirical variance
\begin{equation}
\label{eq:evm-optprob}
    \hat \varphi_a  = \argmin_{\varphi_a \in \mathcal{F}} 
\frac{1}{n} \sum_{i=1}^n \left[ \frac{R_i}{\hat \kappa (\bm Z_i)}
\hat\chi_a(\bm O_i; \hat m_a, \hat g_a)
 - 
\left(
\frac{R_i}{\hat \kappa (\bm Z_i)} - 1
\right) \varphi_a(\bm Z_i)\right]^2.
\end{equation}
When the function class $\mathcal{F}$ contains the true regression function $\varphi_a$ and $m_a$ and $g_a$ are consistently estimated, solving (\ref{eq:evm-optprob}) is equivalent to minimizing the mean squared error $||\hat \varphi_a - \varphi_a||$ asymptotically.
To leverage existing regression software, one can equivalently solve (\ref{eq:evm-optprob}) by fitting a weighted regression of a transformed outcome $\tilde Y = \left((R_i/\hat \kappa (\bm Z_i)) - 1 \right)^{-1}(R_i/\hat \kappa (\bm Z_i))
\hat\chi_a(\bm O_i; \hat m_a, \hat e_a)$ on $\bZ$, with weights $\left((R_i/\hat \kappa (\bm Z_i)) - 1 \right)^2$. We provide further information on implementation of the EEM procedure, and a publicly available software implementation, in the Supplementary Materials. 

\subsection{Optimal Ensembles}
\label{sec:wgtd-avg}

For specific applications, determining which of $\hat \psi_a^\text{OS,1}$ and $\hat \psi_a^\text{OS,2}$ will exhibit stronger finite sample performance \textit{a priori} is an intractable task. This lack of general consensus presents a significant challenge, as researchers will typically prefer to leverage the estimator with greater efficiency, and choosing an estimator based on \textit{post hoc} efficiency checks without adjustment invalidates statistical inference (\citealt{berk2013valid}).
To address this difficulty, we propose weighted averages of the two one-step estimators whose weights are chosen to minimize finite sample variance.
Specifically, we consider weighted averages of the form
$\hat \psi_a^\text{OS,W} = w\hat \psi_a^\text{OS,1} + (1-w)\hat \psi_a^\text{OS,2}, w \in [0,1],$
where we propose setting
\begin{equation}
\label{eq:wgtd-avg-wgts}
    w = \frac{\widehat{\text{Var}}(\hat \psi_a^\text{OS,2}) - \widehat{\text{Cov}}(\hat \psi_a^\text{OS,1},\hat \psi_a^\text{OS,2}) + \delta \hat V}{\hat V - 2\widehat{\text{Cov}}(\hat \psi_a^\text{OS,1},\hat \psi_a^\text{OS,2}) + 2\delta\hat V}.
\end{equation}
Above, $\hat V = \widehat{\text{Var}}(\hat \psi_a^\text{OS,1}) + \widehat{\text{Var}}(\hat \psi_a^\text{OS,2})$ and $\delta>0$ is a small, pre-specified constant. 
Analogues of $\hat \psi_a^\text{OS,W}$ have been explored in data fusion \citep{karlsson2024robust} and high-dimensional \citep{antonelli2020averaging} settings, where in our setting we must account for the fact that both one-step estimators will be asymptotically equivalent under the conditions outlined in Theorems \ref{dr-theorem} and \ref{thm:dr-theorem-alt}.
To this end, the term $\delta \hat V$ is introduced to ensure the weights remain well-defined asymptotically, as $\hat \psi_a^\text{OS,1}$ and $\hat \psi_a^\text{OS,2}$ asymptotically possess the same efficient influence curve. Through the above construction, $\hat \psi_a^\text{OS,W}$ is expected to exhibit finite-sample variance no larger than the lower of the two one-step estimators, while maintaining the same asymptotic distribution as $\hat \psi_a^\text{OS,1}$ and $\hat \psi_a^\text{OS,2}$.
We formalize these notions through the following theorem:
\begin{theorem}
\label{thm:wgtd-avg-dist}
Suppose the conditions of Theorems \ref{dr-theorem} and \ref{thm:dr-theorem-alt} are satisfied so that $\hat \psi_a^\text{OS,1}$ and $\hat \psi_a^\text{OS,2}$ are both RAL for $\psi_a$. Then, when $w$ is determined as in \eqref{eq:wgtd-avg-wgts}, 
\[
\sqrt n (\hat \psi_a^\text{OS,W} - \psi_a) \rightarrow N(0, \sigma_a^2),
\]
where $\sigma^2_a = \E_\Pd[\phi_a(\bm O,\Pd)]^2 = \E_\Pd[\phi_a^\text{ALT}(\bm O,\Pd)]^2.$
\end{theorem}
In practice, we recommend setting $\delta$ to a small positive constant to prevent the term $\delta \hat V$ from distorting the estimation of $w$, recalling the sole function of $\delta \hat V$ is to ensure $\hat \psi_a^\text{OS,W}$ remains well-defined asymptotically. Further details are provided in the Supplementary Materials. In the coming section, we explore the performance of $\hat \psi_a^\text{OS,W}$, as well as our other proposed estimators, through a simulation study.

\section{Simulation Study}
\label{simmy}

\subsection{Setup}

To investigate the performance of the one-step estimators in finite-sample settings, we conducted a set of numerical experiments. Specifically, we generated data according to the following process:
\begin{align*}
    \bX &= (X_1, X_2, X_3)  \sim \text{Uniform}(0, 1) & \text{(Covariates)}
    \\
    A | \bX &\sim \text{Bernoulli}( \ \text{expit}(\bX \bm \delta) \ ) & \text{(Treatment)} 
    \\
    A^* &= A \text{ w.p. } 0.8, \ \ 1-A \text{ w.p. } 0.2 & \text{(Treatment measurement)} 
    \\ 
    Y &= \bX \bm \beta + \tau A + A \bm X \bm \gamma + \varepsilon, \ \ \varepsilon \sim N(0,X_1+X_2+X_3) & \text{(Outcome)}  \\
    Y^* &= Y + \bX \bm \nu + v, \ \ v \sim N(0,1) & \text{(Outcome measurement)}  
    \\
    R | \bX, A^*, Y^* &\sim \text{Bernoulli}\left(\rho \times \frac{\text{expit}(\bm Z \bm \theta)}{\E_\Pd[\text{expit}(\bm Z \bm \theta)]}\right) & \text{(Validation sampling)}
\end{align*}
where $\rho = \PP(R=1)$ controls the relative size of the phase-two sub-sample. 
\\ \\
Across our simulation experiments, we altered (1) the phase-one sample size $n$, and (2) the share of the phase-one sample selected into phase-two, $\rho$. Notably, the coefficients $\bm \theta$ are selected in a manner which over-samples observations with larger values of $A^*$ and $Y^*$ for validation.
\\ \\
Our primary focus centered around the performance of $\hat \psi_a^\text{PI,1}$,  $\hat \psi_a^\text{OS,1}$, $\hat \psi_a^\text{OS,W}$, and two versions of $\hat \psi_a^\text{OS,2}$; one implemented with EEM, and one without. We constructed $\hat \psi_a^\text{OS,W}$ as a weighted average of $\hat \psi_a^\text{OS,1}$ and EEM $\hat \psi_a^\text{OS,2}$, choosing weights according to (\ref{eq:wgtd-avg-wgts}).  Along with the one-step and plug-in estimators, we also considered an oracle estimator where one has access to $A$ and $Y$ for the entire dataset, estimating $\psi$ with augmented inverse probability weighting (AIPW), and a naive estimator where one ignores measurement error and estimates $\psi$ with AIPW by using $Y^*$ and $A^*$ in place of $Y$ and $A$. All nuisance models were estimated with a super learner (\citealt{van2007super}) ensemble model, and the Approach 2 one-step estimators are implemented with the open-source \texttt{drcmd} R package available on GitHub at \texttt{https://github.com/keithbarnatchez/drcmd}. The sampling probabilities $\kappa(\bZ)$ are treated as known, consistent with typical two-phase sampling studies. Full details on the simulation study and super learner libraries used are provided in the Supplementary Materials.

\subsection{Results}

\begin{figure}[h!]
    \centering
    \includegraphics[scale=0.85]{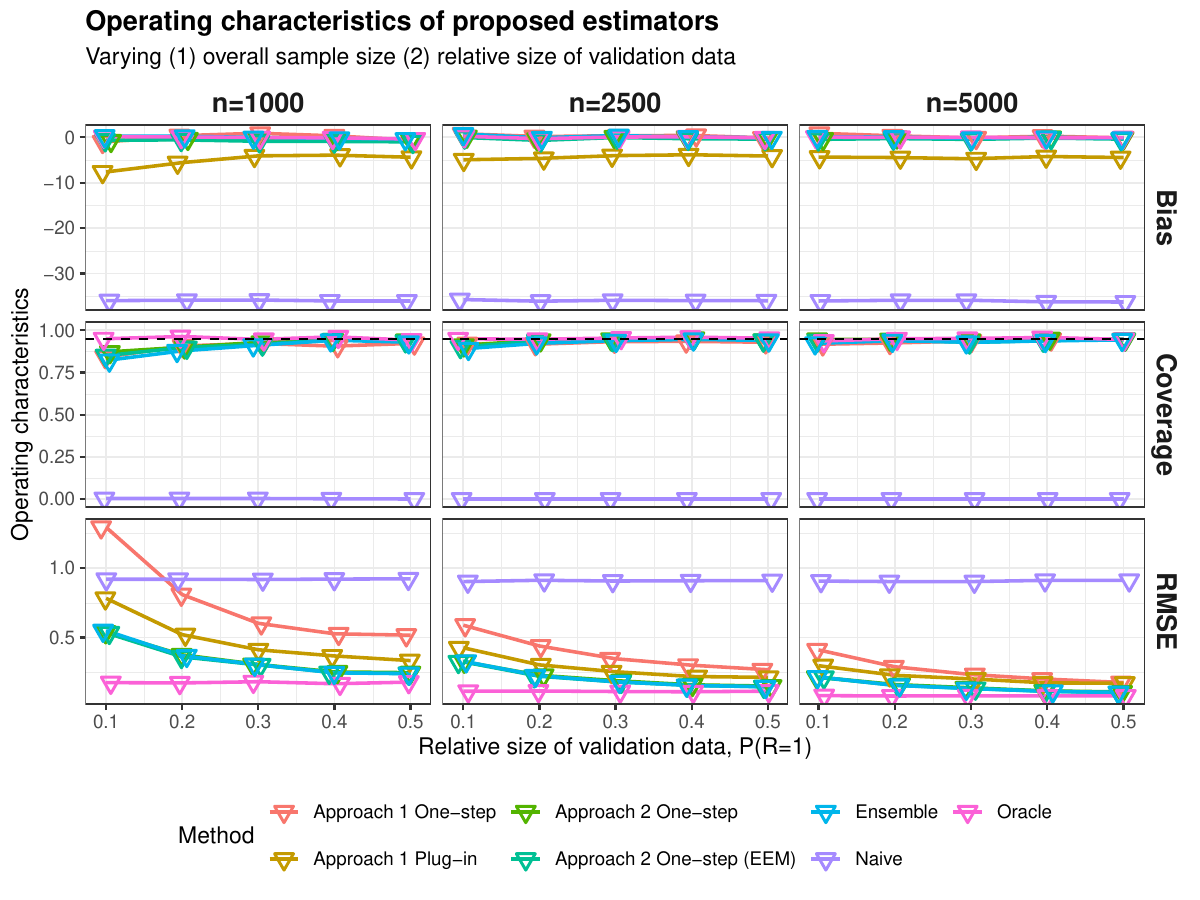}
    \caption{Results of the simulation experiments. For each combination of $n$ and $\rho=\PP(R=1)$ considered, we display the percent bias, root mean squared error (RMSE), and 95\% confidence interval coverage rate of each treatment effect  estimator. RMSE estimates are obtained over $2500$ simulation iterations. A table displaying numerical results is provided in the Supplementary Materials. Random $x$-axis jitter is added to all points.}
    \label{fig:sim-grid}
\end{figure}

Figure~\ref{fig:sim-grid} displays the results of the simulation experiments. 
Across all scenarios considered the naive estimator is severely biased, demonstrating the need to adjust for measurement error, while the Approach 1 plug-in estimator exhibits plug-in bias attributable to the flexible methods used to fit its underlying nuisance functions. 
The remaining estimators are all approximately unbiased. 
Outside of the smallest sample and validation size considered, all proposed one-step estimators attain approximately nominal coverage of 95\% confidence intervals based on empirical asymptotic variance approximations.
\\ \\
Considering the relative efficiencies of our proposed estimators, we note that the Approach 1 one-step estimator is markedly inefficient relative to the Approach 2 one-step estimator.  We reiterate that unstable weights can lead to highly variable Approach 1 one-step estimates. Meanwhile, both versions of the Approach 2 one-step estimator perform similarly, suggesting in our explored setting that sufficiently accurate estimation of $\varphi_a(\bm Z)$ is attainable in small sample sizes.
Most critically, we note that the ensemble estimator is among the set of most efficient estimators across all settings considered. This finding demonstrates the utility of the ensemble estimator in finite-sample settings, where researchers will not know in advance which of $\hat \psi_a^\text{OS,1}$ and $\hat \psi_a^\text{OS,2}$ will exhibit lower variance. 

\section{Data Application}
\label{data-application}

\begin{figure}[h!]
    \centering
    \includegraphics[scale=0.8]{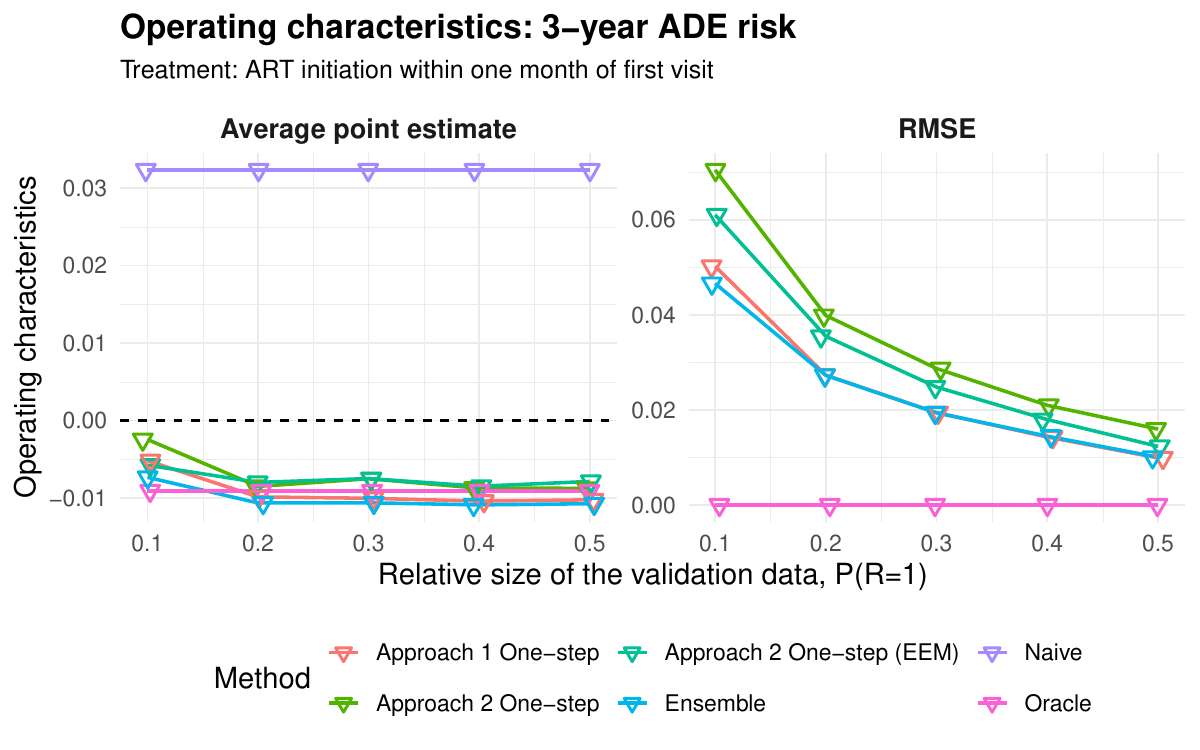}
    \caption{Results of the VCCC data application. The upper panel displays average point estimates across 1,000 iterations at each phase-two relative size considered. The lower panel displays the empirical standard error of each estimator. Random $x$-axis jitter is added to all points.} 
    \label{fig:vccc}
\end{figure}

To study the performance of our proposed methods in a real-world setting, we applied our proposed one-step estimators to EHR data from the Vanderbilt Comprehensive Care Clinic (VCCC), an outpatient clinic providing care for people living with HIV (PLHIV). Throughout each patient's time receiving care from the VCCC, clinical data relevant to the patient's experience were recorded at each visit. Data were also collected on numerous baseline characteristics for each patient, denoted by $\bX_i$, and are outlined  
in Table \ref{tab:vccc-variables}. 
\begin{table}[h!]
    \centering
    \begin{tabular}{ll}
    \toprule
      Variable   & Component  \\
    \midrule
    ADE within 3 years of first visit     & Outcome, $Y$ \\
    \midrule
    Initiated ART within 1 month of first visit & Treatment, $A$ \\
    \midrule
    Sex & Covariate, $\bX$ \\
    Man who has sex with men (MSM) indicator & Covariate, $\bX$ \\
    Injection drug use & Covariate, $\bX$ \\
    Race & Covariate, $\bX$ \\ 
    Ethnicity & Covariate, $\bX$ \\ 
    Age at first visit (years) & Covariate, $\bX$ \\
    Baseline CD4 count (cells/mm$^3$) & Covariate, $\bX$ \\
    \bottomrule
    \end{tabular}
    \caption{VCCC data variables. Note that, as a result of the validation procedure, there are both gold-standard and error-prone versions of $Y$ and $A$.}
    \label{tab:vccc-variables}
\end{table}
A team of researchers validated the charts of all patients in the VCCC EHR database, effectively yielding an initial unvalidated phase-one dataset and a validated phase-two dataset containing gold-standard measurements for all individuals. The validation process revealed a number of substantial errors in key clinical variables, including date of antiretroviral therapy (ART) initiation and occurrence of AIDS-defining events (ADEs). The availability of a full phase-two dataset provides an opportunity to investigate the performance of our proposed estimators over varying relative sizes of phase-two data, allowing us to conduct plasmode simulations in which we control the selection mechanism into the phase-two sample. The VCCC data has been used in numerous studies of measurement error-correction methods~\citep[see, e.g.,][]{oh2021raking, giganti2020accounting, amorim2021two, barnatchez2024flexible}.
\\[1.3em]
For this analysis, we aimed to estimate the average causal effect of early ART initiation ($A$)---defined as starting ART within 1 month of one's initial visit to the VCCC---on the 3-year post-baseline risk of suffering an ADE ($Y$) among patients with no history of ART use prior to initiating care with the VCCC. The initial EHR-derived indicators for early ART ($A^*$) and ADE incidence ($Y^*$) were considerably error prone. Using the validated measurements as gold standards, the misclassification rates of early ART and 3-year ADE incidence were 4.1\% and 12.5\%, respectively.
Following a common exclusion criterion in studies including PLHIV, we excluded individuals who had initiated ART prior to enrollment or suffered an ADE prior to enrollment,  leaving 1,310 study participants.
We considered a grid of relative phase-two sample sizes $\rho = \{0.1,0.2,\ldots,0.5\}$. At each validation size, we simulated 1{,}000 phase-two subsamples sampled without replacement from the original, fully-validated phase-two dataset. At each size considered, we drew validation samples according to $\PP(R_i=1) = \rho \cdot \frac{1 + 0.5A_i^* + Y_i^*}{\frac{1}{n}\sum_{k=1}^n (1+0.5A_k^* + Y_k^*)}$, effectively over-sampling subjects with $A^*=1$ and $Y^*=1$ such that for each $\rho$, $\E[R]=\rho$.
For each simulated dataset, we implemented (1) the Approach 1 and Approach 2 one-step estimators, (2) an oracle AIPW estimator that has access to $Y$ and $A$ for all observations, (3) a naive AIPW estimator that ignores measurement error, using $Y^*$ and $A^*$ for all subjects, (4) a modified Approach 2 one-step estimator that estimates $\varphi_a(\bZ)$ through the EEM procedure outlined in Section \ref{extensions}, and (5) the ensemble estimator presented in Section \ref{sec:wgtd-avg}. All nuisance models were fit with a super learner ensemble \citep{polley2024superlearner} that included generalized linear models with and without interactions \citep{mccullagh1989generalized} and generalized additive models \citep{hastie1986generalized}.
\\[1.3em]
Figure \ref{fig:vccc} displays the main results of our analysis. At each phase-two sample size, we report the average point estimate and RMSE of each method.
The naive estimator exhibits substantial bias, with an average point estimate of 0.0394 suggesting that early ART initiation \textit{increases} the risk of suffering an ADE. The oracle estimate of -0.0112 implies a modest beneficial effect of early initiation of ART, consistent with current scientific consensus on effective treatments for HIV (\citealt{insight2015initiation}). Treating the oracle estimate as the truth, the Approach 1 and both Approach 2 one-step estimators exhibit a degree of small-sample bias---where flexible nuisance models the oracle method is able to make use of are of less utility---that decays for modest phase-two sample sizes. 
\\[1.3em]
Focusing on efficiency, notice the Approach 2 one-step estimator, which estimates $\varphi_a$ through conventional means, is notably inefficient at smaller phase-two sample sizes. This is consistent with our conjecture that poor estimation of $\varphi_a$ will tend to reduce efficiency. The Approach 2 one-step estimator which estimates $\varphi_a$ through empirical efficiency maximization exhibits markedly improved efficiency for all phase-two sample sizes considered. In this setting, the Approach 1 one-step estimator is relatively more efficient, suggesting extreme weights play a lesser role in estimation relative to the simulation study. We again find that the ensemble estimator performs as well or better than all estimators in terms of RMSE across all sample sizes considered.

\vspace{-1em}
\section{Discussion}
\label{discussion}

Measurement error poses a significant challenge to performing causal inference with EHR data. Two-phase sampling designs provide a means to address the bias induced by measurement error, and more general problems where crucial variables are expensive to measure. In this paper, we have presented novel  semi-parametric  efficient estimators for our specific measurement error problem of interest, while providing insight into the general problem of causal inference under two-phase sampling designs. We explicitly linked two general approaches to constructing semi-parametric efficient estimators in two-phase sampling designs, noting that in practice researchers tend to take one approach without reference to the other. We identified unique factors that can result in poor finite sample performance for either approach and presented modifications to the Approach 2 estimator that make use of the empirical efficiency maximization framework. In our applied data example, we demonstrated that our proposed modifications can yield substantial variance reductions without inducing bias. Critically, our proposed ensemble estimator circumvents the need to choose a single approach \textit{a priori},  producing an estimator with optimal finite sample behavior.
\\[1.3em]
Along with providing causal inference practitioners with methods for addressing joint outcome and exposure measurement error, our findings have several implications for the analysis of two-phase sampling data for observational causal inference. For a fixed causal estimand, the efficient influence curve representation (\ref{eq:alt-eif}) used in Approach 2 provides a general means to construct semi-parametric efficient estimators across different two-phase sampling problems. Specifically, one-step estimators derived through Approach 2 will 
take the same form, in terms of the underlying nuisance functions, across two-phase sampling studies with different sets of partially missing variables.
Our work suggests that estimating the pseudo-outcome regression $\varphi_a$ through empirical efficiency maximization provides a straightforward and effective means to address finite-sample inefficiency that can arise through outright estimation of $\varphi_a$. 
Conversely, the forms of doubly-robust estimators derived from Approach 1 can vary substantially across different instances of two-phase sampling problems, as the specific form of the efficient influence curve in one two-phase sampling problem may significantly differ from the efficient influence curve corresponding to a separate two-phase sampling problem with different sets of variables subject to missingness. In settings where one is able to derive Approach 1 estimators, our proposed weighted estimator provides a means to guard against different sources of finite-sample instability.
\\[1.3em]
There are multiple avenues for future work. While our findings suggest that the Approach 2 estimators both (1) enjoy less stringent consistency conditions relative to Approach 1, and (2) can be modified in a straightforward manner to achieve desirable finite-sample performance, similar modifications to improve the finite sample stability of the Approach 1 one-step estimators would be valuable. Further, while we restrict attention to \textit{fixed} two-phase designs, it would be valuable to extend our proposed methods to accommodate \textit{adaptive} designs \citep{wang2023maximin} that aim to select validation samples in a manner that optimizes the asymptotic efficiency of the resulting estimator.  

\subsection*{Funding and Acknowledgements}
This work was funded by the National Institutes of Health (NIH) grants T32AI007358, K01ES032458, R37AI131771, P30AI110527 and P30AI060354-21.

\subsubsection*{Data Availability Statement}

The data utilized in this study were obtained from Vanderbilt University Medical Center under a data use agreement (DUA) and are not publicly available. Access to the data is subject to approval by Vanderbilt University Medical Center and may be requested  through direct inquiry to the institution.

\bibliographystyle{apalike}
\bibliography{sources}

\clearpage


\appendix

\setcounter{equation}{0}
\setcounter{table}{0}
\renewcommand{\theequation}{A\arabic{equation}}
\renewcommand{\thetable}{A\arabic{table}}
\renewcommand{\thefigure}{A\arabic{figure}}

\section{Proofs}

\subsection{Proof of Theorem \ref{id-thm}}

Let $\bm W = (Y^*, A^*)$ and $Z = (\bW, \bX)$. First, notice that Assumptions \ref{sutva1}-\ref{val-positivity6} imply
\begin{align*}
\E(Y(a)) &= \E_{\bm X} \E(Y(a)|\bm X) \\
&= \E_{\bm X} \E(Y|\bm X,A=a) \\
&= \E_{\bW|\bm X,A=a} \ [\E_{\bm X} \E (Y|\bm X, \bm W,A=a)] \\ 
&= \E_{\bW|\bm X,A=a} \ [\E_{\bm X} \E (Y|\bm X, \bm W,A=a,R=1)] 
\end{align*}
The first and second lines hold by iterated expectations and unconfoundedness and consistency. The fourth line holds by the independence condition 
 $Y \indep R | \bX, \bW, A=a$, which holds since the MAR assumption $(Y,A) \indep R | \bm X, \bm W$ implies $Y \indep R | \bm X, \bm W, A$. 
\\ \\
Note that the above expression remains unidentified, since the conditional distribution $\bW|\bX, A=a$ is not identified due to the missingness of $A$. We cannot further condition on $R$, where $A$ is observed, since $A^*$ and $Y^*$ are direct causes of $R$. To address this, we follow \cite{kennedy2020efficient} and note under Assumptions \ref{sutva1}-\ref{val-positivity6} we can write
\begin{align*}
    \psi_a \eqdef \E[Y(a)] &= \E_{\bW|\bm X, A} \ [\E_{\bm X} \E (Y|\bm X, \bW,A=a,R=1)] \\
    &=
    \int \int \int  y p(y|x,w,a,r=1)p(w|x,a)p(x) dy \ dw \ dx \\
    &=
    \int \int \left( \int  y p(y|x,w,a,r=1) dy \right ) \frac{p(a|x,w)p(w|x)p(x)}{p(a|x)p(x)}p(x) \ dw \ dx  \\
    &=
    \int \int \mu_a(z) \frac{p(a|x,w,r=1)p(w|x)}{\int p(a|x,w',r=1)p(w'|x)dw'}p(x) \ dw \ dx \\
    &=
    \int \int \mu_a(z) \frac{\lambda_a(z)p(w|x)}{\int \lambda_a(z)p(w'|x)dw'}p(x) \ dw \ dx \\
    &=
    \int \frac{1}{\pi_a(x)}\int \mu_a(z) \lambda_a(z)p(w|x)p(x) \ dw \ dx \\
    &= 
    \int \frac{\eta_a(x)}{\pi_a(x)}p(x)dx \\
    &=
    \E \left[ \frac{\eta_a(\bX)}{\pi_a(\bX)} \right] = \E\left[ \frac{\E(\mu_a(\bZ) \cdot \lambda_a(\bZ) | \bX)}{\E(\lambda_a(\bZ) | \bX)} \right]
\end{align*}
Above, line 3 holds by Bayes' Rule, and line 4 by iterated expectations and the missing at random assumptions. The remaining lines hold by rearranging and recalling the earlier nuisance function definitions
\begin{align*}
& \lambda_a(\bZ) = \lambda_a(\bX,\bW) = \PP(A=a|\bX,\bW,R=1) & \mu_a(\bZ) = \E[Y|\bX,A=a,\bW,R=1] \\
& \pi_a(\bX) = \E[\lambda_a(\bZ) | \bX] & \eta_a(\bX) = \E[\mu_a(\bZ) \cdot \lambda_a(\bZ) | \bX)].
\end{align*}

\subsection{Proof of Theorem \ref{if-thm}} 

Throughout, let 
$$
\chi_a(\bm  O,\Pd^\text{C})
$$
be the complete data influence curve, and recall $\bm O = (Y, A, \bX), \ \bm W = (Y^*,A^*), \ \ \bZ = (\bX,\bW)$. Let
$
\phi_a(\bm Z,\Pd) 
$
be the observed data influence curve. Following \cite{hou2025efficient}, who take the approach developed by \cite{robins1994estimation}, we leverage the following mapping between the observed data influence curve and complete data influence curve: 
$$
\phi_a(\bZ,\Pd) = \E(\chi_a(\bm O,\Pd^\text{C}) | \bZ) + \frac{R}{\PP(R=1|\bZ)}\left(\chi_a(\bm O,\Pd^\text{C})  - \E(\chi_a(\bm O,\Pd^\text{C}) | R=1, \bm Z) \right),
$$
First, note that the full data statistical estimand---the estimand we would target under access to the complete data---is $\E[\E[Y|A=a,\bX]]$, with corresponding influence curve
\begin{align*}
\chi_a(\bm O,\Pd^\text{C}) &= \frac{I(A=a)}{\PP(A=a|\bX)}(Y-\E(Y|A=a,\bX)) + \E(Y|A=a,\bX) - \psi_a \\
&=
\frac{I(A=a)}{\pi_a(\bX)}\left(Y-\frac{\eta_a(\bX)}{\pi_a(\bX)}\right) + \frac{\eta_a(\bX)}{\pi_a(\bX)} - \psi_a,
\end{align*}
where the validity of the second line was established through the proof of Theorem \ref{id-thm}. Thus, finding $\phi_a(\bX,\Pd)$ amounts to expanding $\E[\chi_a(\bm O,\Pd^\text{C})  | \bZ].$ Notice
\begin{align*}
    \E(\chi_a(\bm O,\Pd^\text{C}) |\bZ) &= \E\left[\frac{I(A=a)}{\pi_a(\bX)}\left(Y-\frac{\eta_a(\bX)}{\pi_a(\bX)}\right) + \frac{\eta_a(\bX)}{\pi_a(\bX)} - \psi_a \bigg| \bZ \right] \\
    &=
    \E\left(\frac{I(A=a)Y}{\pi_a(\bX)} \bigg| \bZ \right) -
     \E\left( \frac{I(A=a)\eta_a(\bX)}{\pi_a(\bX)^2} \bigg| \bZ \right) +
      \E\left( \frac{\eta_a(\bX)}{\pi_a(\bX)} \bigg| \bZ \right) -
       \E\left( \psi_a | \bZ \right) \\
&=
\frac{\lambda_a(\bZ) \mu_a(\bZ)}{\pi_a(\bX)}
- \frac{\lambda_a(\bX)\eta_a(\bX)}{\pi_a(\bX)^2} + 
\frac{\eta_a(\bX)}{\pi_a(\bX)} - \psi_a
\end{align*}
Above, the third line holds via the conditional independence $Y(a) \indep A | \bX$ and through iterated expectations. Plugging this expression back in, we have:
{\footnotesize
\begin{align*}
&\phi_a(\bm O,\Pd)  \\
&= \frac{\lambda_a(\bZ) \mu_a(\bZ)}{\pi_a(\bX)}
- \frac{\lambda_a(\bZ)\eta_a(\bX)}{\pi_a(\bX)^2} + 
\frac{\eta_a(\bX)}{\pi_a(\bX)} - \psi_a \\
&+
 \frac{R}{\PP(R=1|\bZ)}\bigg[\frac{I(A=a)}{\pi_a(\bX)}\left(Y-\frac{\eta_a(\bX)}{\pi_a(\bX)}\right) + \frac{\eta_a(\bX)}{\pi_a(\bX)} - \psi_a  -
\left(\frac{\lambda_a(\bZ) \mu_a(\bZ)}{\pi_a(\bX)}
- \frac{\lambda_a(\bZ)\eta_a(\bX)}{\pi_a(\bX)^2} + 
\frac{\eta_a(\bX)}{\pi_a(\bX)} - \psi_a\right) \bigg] \\
&=
\frac{\lambda_a(\bZ) \mu_a(\bZ)}{\pi_a(\bX)}
- \frac{\lambda_a(\bZ)\eta_a(\bX)}{\pi_a(\bX)^2} + 
\frac{\eta_a(\bX)}{\pi_a(\bX)} - \psi_a \\
&+
 \frac{R}{\PP(R=1|\bZ)}\bigg[\frac{I(A=a)}{\pi_a(\bX)}\left(Y-\frac{\eta_a(\bX)}{\pi_a(\bX)}\right)   - \left(
\frac{\lambda_a(\bZ) \mu_a(\bZ)}{\pi_a(\bX)}
- \frac{\lambda_a(\bZ)\eta_a(\bX)}{\pi_a(\bX)^2} 
 \right) \bigg]
\end{align*}
}
Rearranging, we have
\begin{align*}
&\phi_a(\bm O,\Pd) = \frac{\eta_a(\bX)}{\pi_a(\bX)} - \psi_a \\
&+ \frac{\lambda_a(\bZ)}{\pi_a(\bX)} \left( \mu_a(\bZ) - \eta_a(\bX)/\pi_a(\bX) \right)
\\
&+
 \frac{R I(A=a)}{\PP(R=1|\bZ)\pi_a(\bX)}\left(Y-\frac{\eta_a(\bX)}{\pi_a(\bX)}\right) \\
 &- 
\frac{R\lambda_a(\bZ) }{\PP(R=1|\bZ)\pi_a(\bX)}\left(
\mu_a(\bZ)
- \frac{\eta_a(\bX)}{\pi_a(\bX)}
 \right),
\end{align*}
as desired. Further note that this final display additionally proves Proposition \ref{prop:eif-prop}. 
We briefly note  that one could alternatively derive $\phi_a(\bm Z,\Pd)$ through differentiating 
\[
\E_{\Pd_\varepsilon}\left[ \frac{\eta_{a,\varepsilon}(\bX)}{\pi_{a,\varepsilon}(\bX)} \right],
\]
where $\Pd_\varepsilon$ is a generic parametric submodel for $\Pd$, recalling that $\phi_a(\bm O,\Pd)$ will be a mean-zero random variable which satisfies
\[
\frac{d}{d\varepsilon}
\E_{\Pd_\varepsilon}\left[ \frac{\eta_{a,\varepsilon}(\bX)}{\pi_{a,\varepsilon}(\bX)} \right] \bigg|_{\varepsilon=0} = 
\E_\Pd[\phi_a(\bm O,\Pd)u(\bm O)],
\]
where $u(\bm O)$ is the parametric submodel score evaluated at $\varepsilon=0$. One will arrive at the same influence curve.

\subsection{Proof of Theorem \ref{dr-theorem}}

Our goal is to show that 

$$
\E[\hat \psi_a^\text{OS,1} - \psi_a]= o_\Pd \left( \frac{1}{\sqrt n} \right) +
O_\Pd \left((||\hat \eta_a - \eta_a ||  + || \hat \pi_a - \pi_a ||) \cdot || \hat \pi_a - \pi_a || + || \hat \varphi_a - \varphi_a || \cdot || \hat \kappa - \kappa || \right)
$$
Following standard approaches \citep{kennedy2024semiparametric}, we consider the expansion
\begin{align}
 \E[\hat \psi_a^\text{OS,1} - \psi_a] &= (\Pd_n - \Pd)\phi_a + (\Pd_n - \Pd)(\hat \phi_a - \phi_a) + \Pd(\hat \phi_a - \phi_a)  \nonumber \\
 &= T_1 + T_2 + T_3 \label{eq:dr-proof-expansion1}
\end{align}
where we will occasionally omit the observational and distributional arguments from the influence function $\phi_a$ for brevity.
Above, the first term $T_1$ is $O_\Pd(1/\sqrt n)$ by the Central Limit Theorem, while the second term  can be controlled by analogous arguments made in the proof of Theorem \ref{dr-theorem}. $T_3$ can be studied by using the result from \cite{levis2024double} that we can write
\begin{align}
\E[\hat \psi_a^\text{OS,1} - \psi_a] &= 
\underbrace{\psi_a^\text{PI,1}(\hat{\Pd}^\text{C}) 
+ \E[\chi_a(O,\hat{\Pd}^\text{C})] - \psi_a(\Pd^\text{C})}_\text{full data bias}  \nonumber \\
&+ \E\left[ \left( 1 - \frac{\kappa(\bZ)}{\hat \kappa(\bZ)} \right) \left\{ \frac{\hat \lambda_a(\bm Z)\hat \mu_a(\bm Z)}{\lambda_a(\bm Z) \mu_a(\bm Z)}  - 1\right\} \right] \label{eq:full-data-bias}
\end{align}
We begin by focusing on the first term. Letting $\hat \eta_a/\hat\pi_a = \hat m_a$ and $\hat \pi_a = \hat g_a$, we note that
\begin{align*}
\psi_a^\text{PI,1}(\hat{\Pd}^\text{C}) + \E[\chi_a(\bm O, \hat{\Pd}^\text{C})] - \Psi_a(\Pd^\text{C}) &= 
o_{\Pd}  ( || \hat m_a - m_a || \cdot ||\hat g_a - g_a|| ),
\end{align*}
where the above decomposition is a well-known property of the full-data influence curve (\citealt{bang2005doubly}). Substituting in the equalitites $m_a=\eta_a/\pi_a$, $g_a=\pi_a$ derived in the proof of Theorem \ref{id-thm}, we have that
\begin{align*}
\psi_a^\text{PI,1}(\hat{\Pd}^\text{C}) + \E[\chi_a(\bm O, \hat{\Pd}^\text{C})] - \Psi_a(\Pd^\text{C}) &=
o_{\Pd}  ( ||\hat \eta_a - \eta_a ||  + || \hat \pi_a - \pi_a ||)\cdot ||\hat \pi_a - \pi_a|| ),
\end{align*}
Noting the second term of \eqref{eq:full-data-bias} is $O_\Pd(|| \hat \kappa - \kappa|| \cdot (||\hat \mu_a - \mu_a||+||\hat \lambda_a - \lambda_a|| ) ) $,
and recalling the rates associated with each term in (\ref{eq:dr-proof-expansion1}) yields the desired result.

\subsection{Proof of Theorem \ref{thm:dr-theorem-alt}}

Our goal is to show that 

$$
\E[\hat \psi_a^\text{OS,2} - \psi_a]= o_\Pd \left( \frac{1}{\sqrt n}\right)  + O_\Pd \left( ||\hat m_a - m_a ||  \cdot || \hat g_a - g_a || + || \hat \varphi_a - \varphi_a || \cdot || \hat \kappa - \kappa || \right)
$$

It is well-established that in general one-step estimation problems, one can study $\E[\hat \psi_a^\text{OS,2} - \psi_a]$ by considering the following error decomposition (see, e.g. \citealt{kennedy2024semiparametric}):
\begin{align}
 \E[\hat \psi_a^\text{OS,2} - \psi_a] &= (\Pd_n - \Pd)\psi_a + (\Pd_n - \Pd)(\hat \psi_a - \psi_a) + \Pd(\hat \psi_a - \psi_a)  \nonumber \\
 &= T_1 + T_2 + T_3 \label{eq:dr-proof-expansion}
\end{align}
Above, the first term $T_1$ is $o_\Pd(1/\sqrt n)$ by the Central Limit Theorem, while the second term  is $o_\Pd(1/\sqrt n)$ if $\E[\hat \psi_a - \psi_a] = o_\Pd(1)$ and any one of the following three conditions hold: (i) all nuisance models are trained on a separate held-out sample, (ii) one constructs $\hat \psi_a^\text{OS,1}$ through cross-fitting, or (iii) all nuisance functions are contained within a Donsker class. For simplicity we operate under case (i), though in practice we recommend the use of cross-fitting, through which the same robustness properties will hold. 
\\ \\
The third term $T_3$ is typically referred to as a \textit{second-order remainder} or \textit{conditional bias} term, and requires closer study. While one can manually inspect
$$
 \E[\hat \psi_a^\text{OS,2} - \psi_a] = \hat \psi_a^\text{PI,2}(\hat{\Pd}) + \E[\Phi_a(O,\hat{\Pd})] - \Psi_a(\Pd),
$$
a more straightforward approach is to leverage the bias structure of the full data influence curve. We again leverage Proposition A.4 of \cite{levis2024double}, which shows one can alternatively write $T_3$ as
\begin{align*}
\E[\hat \psi_a^\text{OS,1} - \psi_a] &= 
\underbrace{\psi_a^\text{PI,1}(\hat{\Pd}^\text{C}) +  \E[\chi_a(\bm O, \hat{\Pd}^\text{C})] - \Psi_a(\Pd^\text{C})}_\text{full data bias} \\
&+ \E\left[ \left( 1 - \frac{\kappa(\bZ)}{\hat \kappa(\bZ)} \right) \{\hat \varphi_a(\bZ) - \varphi_a(\bZ)\right]
\end{align*}
In words, the bias of the observed-data one-step estimator can be written as the sum of (1) the bias of the corresponding full-data one-step estimator, and (2) a product of biases between the sampling probabilities and the full-data influence curve projected into the observed data tangent space. 
\\ \\
The second term above is $O_\Pd(||\hat \kappa - \kappa || \cdot ||\hat \varphi_a - \varphi_a ||)$ by Cauchy-Schwartz. We can therefore turn our attention to the first term. Recall from the proof of Theorem \ref{if-thm} that
\begin{align*}
\psi_a^\text{PI,1}(\hat{\Pd}^\text{C}) + \E[\chi_a(\bm O, \hat{\Pd}^\text{C})] - \Psi_a(\Pd^\text{C}) &= m_a(\bW) +  \frac{I(A=a)}{\pi_a(\bW)} (Y-m_a(\bW)) \\
&=
O_{\Pd}  ( || \hat m_a - m_a || \cdot ||\hat \pi_a - \pi_a|| ),
\end{align*}
where the second line is a well-known property of the full data influence curve for the ATE functional $\E[\E(Y|A=a,\bX)]$.
Recalling the rates associated with each term in (\ref{eq:dr-proof-expansion}) yields the desired result.

\subsection{Proof of Theorem \ref{thm:wgtd-avg-dist}}

Under the conditions of Theorems \ref{dr-theorem} and \ref{thm:dr-theorem-alt}, notice that we have
\begin{align*}
    \hat \psi_a^\text{OS,1} - \psi_a &= \sumn \phi_a(\bm O,\Pd) + a_n, \text{ and} \\
       \hat \psi_a^\text{OS,2} - \psi_a &= \sumn \phi_a(\bm O,\Pd) + b_n, 
\end{align*}
where $a_n$ and $b_n$ are both $o_\Pd(1/\sqrt n)$.
Let $\hat \sigma^2_{a,1}$ and $\hat \sigma^2_{a,2}$ denote estimators for $\sigma^2_a$ constructed from Approach 1 and Approach 2. Further let $\hat \rho$ be an estimator for $\text{Cov}(\hat \psi_a^\text{OS,1},\hat \psi_a^\text{OS,2})$, where all 3 estimators are based on their empirical influence curves.
By the asymptotic linearity of the two one-step estimators, we have that $\hat \sigma^2_{a,1} \cp \sigma^2_a$, $\hat \sigma^2_{a,2} \cp \sigma^2_a$, and $\hat \rho_a \cp \rho_a=\sigma^2_a$, where the final equality holds since $\sigma^2_{a,1}=\sigma^2_{a,2}$. Recalling 
\[
\hat w = \frac{\hat \sigma^2_{a,2} - \hat \rho_a + \delta \hat V}{\hat \sigma^2_{a,1} + \hat \sigma^2_{a,2}- 2\hat \rho_a + 2\delta\hat V}, \ \ \delta > 0
\]
notice we will have
\begin{align*}
    \hat w &\cp \frac{ \sigma^2_{a} -  \rho_a + \delta ( \sigma^2_{a} + \sigma^2_{a})}{ \sigma^2_{a} +  \sigma^2_{a}- 2 \rho_a + 2\delta ( \sigma^2_{a} + \sigma^2_{a})} \\
    &=
    \frac{ \delta ( \sigma^2_{a} + \sigma^2_{a})}{ 2\delta ( \sigma^2_{a} + \sigma^2_{a})} \\
    &= \frac{1}{2},
\end{align*}
implying $\hat w = 1/2 + o_\Pd(1)$. 
We briefly note that without the inclusion of the term $\delta \hat V$, the proposed weights would be undefined asymptotically, and that choosing a small value for $\delta$ prevents its inclusion from severely impacting the finite-sample optimality of $\hat w$.
The above result implies
\begin{align*}
    \hat \psi_a^\text{OS,W} - \psi_a &= \hat w  \hat \psi_a^\text{OS,W} + (1-\hat w) \hat \psi_a^\text{OS,2} \\
    &=\sumn \phi_a(\bm O_i,\Pd) + \frac{1}{2}\left( a_n + b_n \right) + o_\Pd(1)(a_n + b_n) \\
    &=
    \sumn \phi_a(\bm O_i,\Pd) + c_n,
\end{align*}
where $c_n = o_\Pd(1/\sqrt n)$, implying $\hat \psi_a^\text{OS,W}$ is asymptotically linear with influence curve $\phi_a(\bm O,\Pd)$, as desired.

\subsection{Identification of $\Pd^\text{C}$ from $\Pd$}

For brevity, let $p(v) = d\Pd(v)/dv$ and  $p^\text{C}(v) = d\Pd^\text{C}(v)/dv$. 
Following the strategy used in \cite{levis2024double}, notice the density of the complete data can be written
\begin{align*}
   p^\text{C}(\bX,A,Y,A^*,Y^*,R) &= p^\text{C}(\bX,A^*,Y^*,RY,RA,(1-R)Y,(1-R)A,R) \\
   &=
p(\bX,A^*,Y^*,RY,RA, R)p^\text{C}(Y,A| \bX, Y^*, A^*, R=0)^{1-R} \\
&=
p(\bX,A^*,Y^*,RY,RA,R)p(Y,A| \bX, Y^*, A^*, R=1)^{1-R} 
\end{align*}
Above, the final line holds by Assumptions \ref{ymcar4} and \ref{val-positivity6}. This final line implies $\Pd^\text{C}$ is identified by $\Pd$, since both component densities are with respect to the observed data distribution.

\clearpage

\section{Empirical efficiency maximization}

In this Section, we provide further justification for our proposed implementation of the empirical efficiency maximization estimator.  

\subsection*{Loss Function Motivation}

To motivate the EEM loss function, consider a quasi-oracle setting where the nuisance functions $m_a, g_a$ and $\kappa$ are known so that $\hat m_a=m_a, \hat g_a=g_a$ and $\hat \kappa=\kappa$. The definition of $\varphi_a(\bm Z) = \E[\chi_a(\bm O_i; m_a,  g_a) | \bm Z, R=1]$ implies that $\varphi_a$ satisfies
\begin{equation}
\label{eq:varphi-loss-conv}
  \varphi_a(\bm Z) = \argmin_{\tilde \varphi_a \in \mathcal{M}} = \E \{ R (\chi_a(\bm O ; m_a, g_a) - \tilde \varphi_a(\bm Z))^2 \}  
\end{equation}

Further recall that
\[
\phi_a^\text{ALT}(\bm O; \Pd) = 
\frac{R_i}{\kappa(\bm Z)}
\chi_a(\bm O_i; m_a,  g_a)
 - 
\left(
\frac{R_i}{\kappa(\bm Z_i)} - 1
\right) \tilde\varphi_a(\bm Z_i),
\]
where $\phi_a^\text{ALT}(\bm O; \Pd)$ is mean zero.
Briefly letting $\phi_a^\text{ALT}(\bm O; \tilde\Pd) = \phi_a^\text{ALT}(\bm O; m_a,g_a,\kappa,\tilde \varphi_a)$ and
recalling $\bm O_i \sim \Pd\in\mathcal{M}$, since $\phi_a^\text{ALT}(\bm O; \Pd)$ is the efficient influence function for $\Pd$ in a nonparameteric model $\mathcal{M}$, $\varphi_a$ additionally satisfies
\begin{equation}
\label{eq:varphi-loss-eem}
    \argmin_{\tilde \varphi_a \in \mathcal{M}}  =\E \left\{ \left(
\frac{R_i}{\kappa(\bm Z)}
\chi_a(\bm O_i; m_a,  g_a)
 - 
\left(
\frac{R_i}{\kappa(\bm Z_i)} - 1
\right) \tilde\varphi_a(\bm Z_i)
\right)^2
\right\} = \E[\phi_a^\text{ALT}(\bm O; \tilde\Pd)^2] 
\end{equation}

That $\varphi_a(\bm Z)$ satisfies both (\ref{eq:varphi-loss-conv}) and (\ref{eq:varphi-loss-eem}) implies both displays can be used as loss function for its estimation. 
In the section below, we propose a straightforward means to minimize solve (\ref{eq:varphi-loss-eem}) with conventional regression software.


\subsection*{Equivalence to Weighted Regression}
Recall our goal is to find $\varphi_a$ satisfying
\begin{align*}
&    \hat \varphi_a  = \argmin_{\varphi_a \in \mathcal{M}} 
\frac{1}{n} \sum_{i=1}^n \left[ \frac{R_i}{\PP(R_i=1|\bm Z_i)}
\hat\chi_a(\bm O_i; \hat m_a, \hat e_a)
 - 
\left(
\frac{R_i}{\PP(R_i=1|\bm Z_i)} - 1
\right) \varphi_a(\bm Z_i)\right]^2.
\end{align*}
Notice for any $i$, 
\begin{align*}
& \left[    \frac{R_i}{\PP(R_i=1|\bm Z_i)}
\hat\chi_a(\bm O_i; \hat m_a, \hat e_a)
 - 
\left(
\frac{R_i}{\PP(R_i=1|\bm Z_i)} - 1
\right) \varphi_a(\bm Z_i)\right]^2 
\\
&= \left(\frac{\frac{R_i}{\PP(R_i=1|\bm Z_i)} - 1}{\frac{R_i}{\PP(R_i=1|\bm Z_i)} - 1}\right)^{2} \left[    \frac{R_i}{\PP(R_i=1|\bm Z_i)}
\hat\chi_a(\bm O_i; \hat m_a, \hat e_a)
 - 
\left(
\frac{R_i}{\PP(R_i=1|\bm Z_i)} - 1
\right) \varphi_a(\bm Z_i)\right]^2 
\\
&=
\left(\frac{R_i}{\PP(R_i=1|\bm Z_i)} - 1\right)^{2}
\left[  \left(\frac{R_i}{\PP(R_i=1|\bm Z_i)} - 1\right)^{-1}  \frac{R_i}{\PP(R_i=1|\bm Z_i)}
\hat\chi_a(\bm O_i; \hat m_a, \hat e_a)
 - 
\varphi_a(\bm Z_i)\right]^2 \\
&=
\left(\frac{R_i}{\PP(R_i=1|\bm Z_i)} - 1\right)^{2}
\left[ \tilde Y_i
 - 
\varphi_a(\bm Z_i)\right]^2
\end{align*}
where $\tilde Y_i \eqdef \left(\frac{R_i}{\PP(R_i=1|\bm Z_i)} - 1\right)^{-1}  \frac{R_i}{\PP(R_i=1|\bm Z_i)}
\hat\chi_a(\bm O_i; \hat m_a, \hat e_a)$, implying we can equivalently choose $\hat \varphi_a$ such that 
\begin{align*}
\hat \varphi_a  = \argmin_{\varphi_a \in \mathcal{M}} 
\frac{1}{n} \sum_{i=1}^n \left(\frac{R_i}{\PP(R_i=1|\bm Z_i)} - 1\right)^{2}
\left[ \tilde Y_i
 - 
\varphi_a(\bm Z_i)\right]^2    
\end{align*}
Notice this simply resembles an empirical loss function of a weighted regression of $\tilde Y$ on $\varphi(\bm Z)$, with regression weights $\left(\frac{R_i}{\PP(R_i=1|\bm Z_i)} - 1\right)^{2}$.

\clearpage

\section{Simulation Details}

In this section, we provide further details on the simulation study carried out in Section \ref{simmy}. 

\subsection*{Reproducibility and Software}

Code for reproducing all results from Section \ref{simmy} can be found at \\ \texttt{https://github.com/keithbarnatchez/me-dep-sampling}. All simulations are performed in R \citep{R}.
\\ \\
Given the generality of the Approach 2 one-step estimation method, we developed the R package \texttt{drcmd} which implements the methods described in Section \ref{sec:approach2} both (i) for the measurement error setting considered in this paper, and (ii) more general missing data settings that rely on missing at random assumptions analogous to those made in Assumptions \ref{ymcar4}-\ref{val-positivity6}. Further details on the package can be found at \texttt{https://github.com/keithbarnatchez/drcmd}. The \texttt{drcmd} package is used to implement all Approach 2 estimators in our simulations, both with and without EEM.

\subsection*{Nuisance Learners}

In implementing the proposed estimators described in Section \ref{sec:methods}, we fit all nuisance functions with a Super Learner ensemble, specifying the libraries \texttt{SL.glm}, \texttt{SL.glm.interaction}, and \texttt{SL.gam}. The sampling probabilities are treated as known in our main exercise. In our additional exercise where we estimate $\kappa$, we use the same libraries specified above.

\subsection*{Simulation Parameters}

The table below outlines the values of parameters specified in Section \ref{simmy}.

\begin{table}[h!]
    \centering
    \begin{tabular}{lll}
    \toprule
Parameter & Description & Value(s)  \\
\midrule 
$\bm \delta$ & Propensity score coefficients & $(-0.1,-0.6,-0.9)$ \\
$\bm \beta$ & Outcome model covariate effects & $(1,2,-2)$ \\
$\tau$ & Constant component of treatment effect & 1 \\
$\bm \gamma$ & Interaction coefficients & $(1,1,1)$ \\
$\bm \nu$ & Outcome measurement error coefficients & $(0.1,-0.1,0.1)$ \\
$\bm \theta$ & Validation sampling coefficients & $(0.6,-0.2,0.8,0.1,-0.3)$ \\
\midrule
$n$ & Sample size & $1000, \ 2500, \ 5000$ \\
$\rho$ & Relative size of phase-two data & $0.1,0.2,0.3,0.4,0.5$ \\
\bottomrule
    \end{tabular}
    \caption{Simulation parameter values. $\rho$ and $n$ vary across simulation scenarios, other parameters remain fixed.}
    \label{sim-params}
\end{table}

\clearpage

\section{Data Application Details}

In this section, we detail the implementation of our data application with the VCCC database. Our procedure closely follows the approach outlined in \cite{barnatchez2024flexible}.
\\ \\
We initialized a grid of validation relative sizes of the form $\rho = \{0.1,0.2,\ldots,0.5\}$. For each point on the grid, we repeated the following procedure 1{,}000 times:
\begin{enumerate}
    \item We obtained a random sample of validated exposure and outcome measurements according to the sampling rule outlined in Section \ref{data-application}. The current point on the relative size grid, $\rho$, is used to normalize the sampling rule so that $\PP(R=1)=\rho$.
    \item Using the randomly sampled phase-two data and remaining unvalidated data, we implement our proposed methods to obtain estimates of $\psi$. We estimate nuisance functions through a Super Learner, specifying libraries \texttt{SL.glm}, \texttt{SL.glm.interaction} and \texttt{SL.gam}.
\end{enumerate}
For each $\rho$, we then average estimates across the 1{,}000 datasets. Additionally, we obtain a single ``oracle" and ``naive" estimate from the full dataset, using the same procedure outlined in Section \ref{simmy}.

\clearpage

\section{Targeted Maximum Likelihood}

While the Approach 2 one-step estimator removes the asymptotic bias incurred by $\hat \psi_a^\text{PI,2}$ through augmentation of the term $\hat \phi_a^\text{ALT}(O_i,\hat \Pd)$, TML proceeds by updating the initial nuisance function estimates in a way that removes the asymptotic bias of the updated plug-in estimator. A TML estimator of $\psi_a$ can be carried out in a two-step procedure originally proposed by \cite{rose2011targeted} for two-phase sampling designs:
\begin{enumerate}
\item Obtain updated sampling probabilities through a logistic regression working model
    $
    \text{logit}(\hat \kappa(\bm Z ; \zeta)) = \text{logit}(\hat \kappa(\bm Z)) + \zeta (\hat \varphi_a(\bm Z)/\hat \kappa(\bm Z)),
    $
     fitting $\zeta$ through maximum likelihood and denoting the updated values by $\hat \kappa^*(\bm Z)$.
    \item Using the updated sampling probabilities, similarly fit a weighted logistic regression working model
    $
    \text{logit}(\hat m_a(\bX;\varepsilon)) = \text{logit}(\hat m_a(\bX)) + 
    \varepsilon \left(  I(A=a)/\hat g_a(\bX) \right)
    $
    with weights $R/\hat \kappa^*(\bm Z)$. Denote the updated values by $\hat m_a^*(\bX)$.
\end{enumerate}
 The above implementation yields an updated plug-in estimator 
 \begin{equation}
     \hat \psi_a^\text{TML-ALT} = \sumn \hat m_a^*(\bX_i),
 \end{equation}
which enjoys the same asymptotic properties as $\hat \psi_a^\text{OS,2}$ under standard regularity conditions; see e.g. \cite{hejazi2021efficient} for further discussion. To provide further intuition, recall that $\hat \psi_a^\text{OS,2}$ removes the plug-in bias of $\hat \psi_a^\text{PI,2}$ through augmentation of the term
\begin{align*}
&\sumn \bigg \{ \frac{R_i}{\kappa(\bZ_i)}\left(\frac{I(A_i=a)}{\hat g_a(\bX_i)}(Y-\hat m_{A_i}(\bX_i)) + \hat m_a(\bX_i)  - \hat \psi_a^\text{PI,2}\right)  \\
& \ \ \ \  \  \ \ \ \ \  \ \ -  \left( \frac{R_i}{\kappa(\bm Z_i)} - 1\right) \hat \varphi_a(\bZ_i) \bigg \}.
\end{align*}
The above TML implementation instead removes this plug-in bias by ensuring $\hat \kappa^*(\bZ)$ and $\hat m_a^*(\bX)$ satisfy
\begin{align*}
&\sumn \bigg \{ \frac{R_i}{\hat \kappa^*(\bZ_i)}\left(\frac{I(A_i=a)}{\hat g_a(\bX_i)}(Y-\hat m^*_{A_i}(\bX_i)) + \hat m^*_a(\bX_i)  - \hat \psi_a^\text{PI,2}\right)  \\
& \ \ \ \  \  \ \ \ \ \  \ \ -  \left( \frac{R_i}{\hat \kappa^*(\bm Z_i)} - 1\right) \hat \varphi_a(\bZ_i) \bigg \} = 0.
\end{align*}
Notably, the pseudo-outcome regression $\hat \varphi_a(\bm Z)$ is only leveraged as a covariate used to adjust the initial sampling probabilities $\hat \kappa(\bm Z)$. Relative to $\hat \psi_a^\text{OS,2}$, where $\hat \varphi_a(\bm Z)$ directly contributes to a sample average, the construction of $\hat \psi_a^\text{TML}$ dampens the impact of individual terms that may otherwise exert a large influence in small samples. The impact of large weights $1/\hat \kappa(\bm Z)$ is similarly mitigated. The notion of TML enjoying greater stability than one-step estimation methods in finite sample settings is well-documented (\cite{stitelman2010collaborative,ellul2024causal}), and suggests $\hat \psi_a^\text{TML-ALT}$ will likely enjoy greater finite sample stability relative to $\hat \psi_a^\text{OS,2}$. 
\\ \\
Though we do not implement $\hat \psi_a^\text{TML,2}$ in our main simulations for brevity, we do provide an implementation in the \texttt{drcmd} R package used to implement $\hat \psi_a^\text{OS,2}$. Further, we consider simulations which include $\hat \psi_a^\text{TML,2}$ in Section \ref{sec:addl-sim}.

\section{Additional Simulation Exercises}
\label{sec:addl-sim}

\subsection{Estimation of sampling probabilities}

\begin{figure}[h!]
    \centering
    \includegraphics[scale=0.7]{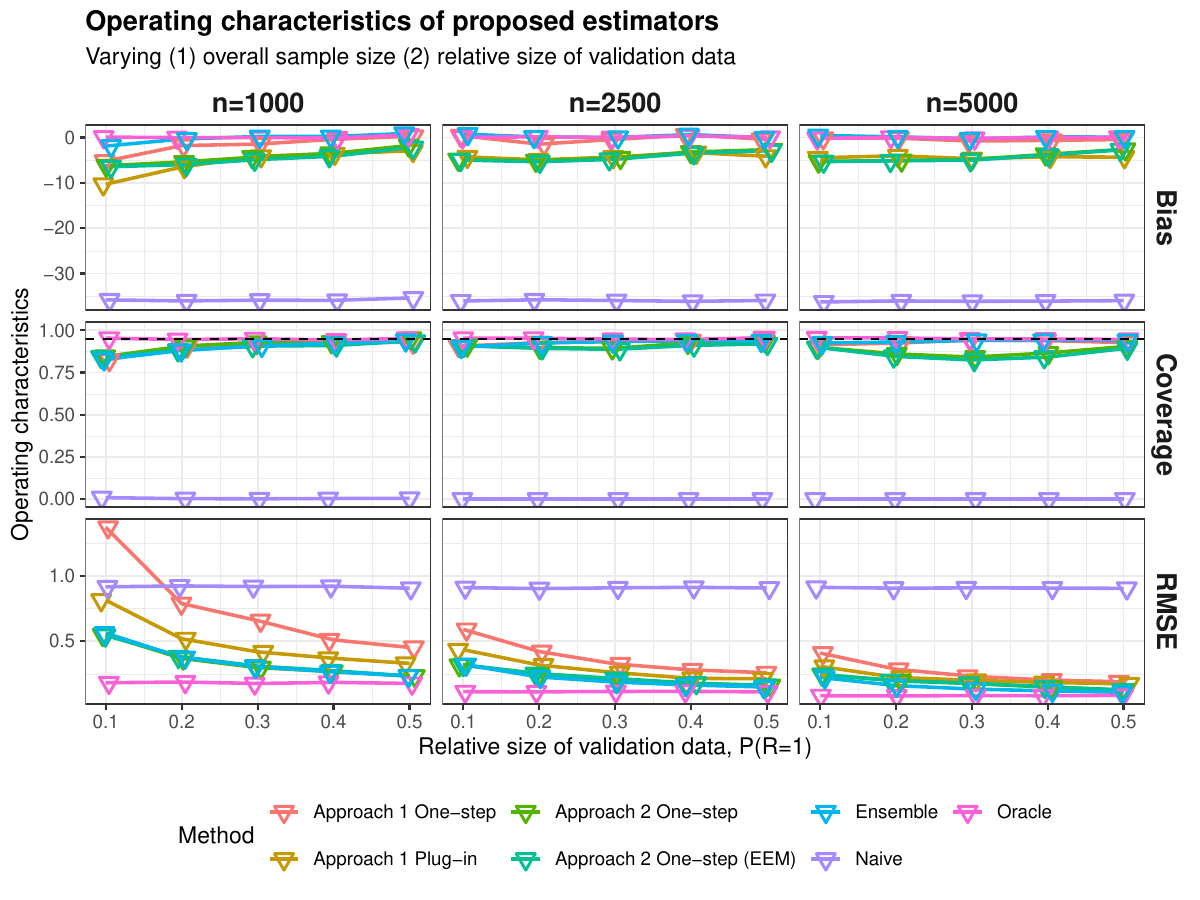}
    \caption{Simulation results, mimicking the setting in Section \ref{simmy}, where the sampling probabilities $\bm \kappa(\bZ)$ are now estimated.}
    \label{fig:ophars-estr}
\end{figure}

While sampling probabilities $\kappa(\bm Z)$ are known in the settings we consider, our methodology easily accommodates settings where $\kappa(\bm Z)$ is estimated. Consideration of such settings may be of interest (i) for settings where the sampling probabilities are not known, or general missing data settings, and (ii) to allow for further efficiency gains, as estimation of $\kappa$ is known to enhance efficiency relative to estimators where true values of  $\kappa$ are used. We estimate $\kappa$ using the same Super Learner ensemble used to estimate all other nuisance functions in our main simulation exercise.
\\ \\
Figure \ref{fig:ophars-estr} displays the results. We see all estimators remain consistent, with similar relative efficiencies to those documented in Section \ref{simmy}.

\subsection{Performance of Targeted Maximum Likelihood Estimator}

\begin{figure}
    \centering
    \includegraphics[scale=0.8]{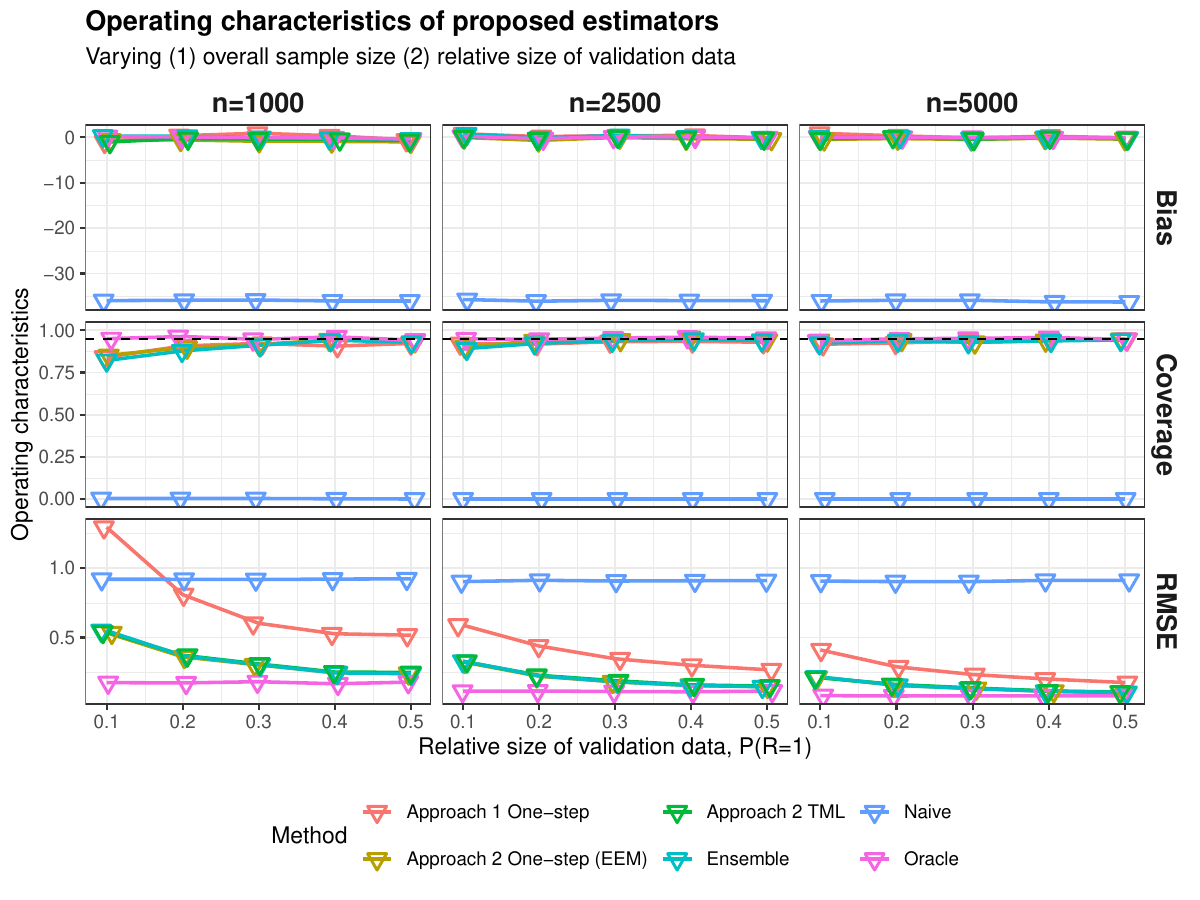}
    \caption{Simulation results under same setting outlined in Section \ref{simmy}, now including $\hat \psi_a^\text{TML,2}$. We remove $\hat \psi_a^\text{PI,1}$ and non-EEM $\hat \psi^\text{OS,2}$ solely for parsimony. }
    \label{fig:sim-tml-incl}
\end{figure}

We additionally consider the performance of $\hat \psi_a^\text{TML}$, otherwise repeating the exercises considered in Section \ref{simmy}. The results can be found in Figure \ref{fig:sim-tml-incl}.  As in Section \ref{simmy}, $\hat \psi_a^\text{OS,W}$ is formed as an ensemble of $\hat \psi_a^\text{OS,1}$ and $\hat \psi_a^\text{OS,2}$. We see that $\hat \psi_a^\text{TML,2}$ performs comparatively to EEM-based $\hat \psi_a^\text{OS,2}$.

\end{document}

%% file: preamble.tex
\usepackage[utf8]{inputenc}

\usepackage{geometry}
\usepackage{natbib}
\usepackage{amsmath}
\usepackage{amssymb}
\usepackage{xcolor}
\usepackage{hyperref}
\usepackage{booktabs}
\usepackage{amsthm}
\usepackage{bm}
\usepackage{pdflscape}
\usepackage{setspace}
\usepackage{graphicx}
\usepackage{subcaption}

\allowdisplaybreaks

\hypersetup{colorlinks=true,linkcolor=blue,filecolor=blue,citecolor=blue}

\usepackage{tikz}
\usepackage[labelfont=bf]{caption}
\usepackage{relsize}
\usepackage{xifthen}
\usepackage{amsmath,amssymb,xcolor}
\usepackage{pgf,tikz}
\usetikzlibrary{arrows,shapes.arrows,shapes.geometric,
shapes.multipart,decorations.pathmorphing,positioning,
swigs}

\newtheorem{theorem}{Theorem}
\newtheorem{assumption}{Assumption}

\numberwithin{intassumption}{assumption}

\expandafter\def\expandafter\normalsize\expandafter{%
    \normalsize%
    \setlength\abovedisplayskip{5pt}%
    \setlength\belowdisplayskip{5pt}%
    \setlength\abovedisplayshortskip{-8pt}%
    \setlength\belowdisplayshortskip{2pt}%
}

\newcommand{\E}{\mathbb{E}}
\newcommand{\PP}{\mathbb{P}}
\newcommand{\bX}{\boldsymbol X}
\newcommand{\bZ}{\boldsymbol Z}

\newcommand{\indep}{\perp \!\!\! \perp}
\newcommand{\cp}{\overset{p}{\rightarrow}}

\newcommand{\eqdef}{\overset{\text{def}}{=}}

\newcommand{\bW}{\boldsymbol W}

\newcommand{\bO}{\bm O}
\newcommand{\Pd}{\mathsf{P}}

\DeclareMathOperator*{\argmin}{arg\,min}

\providecommand{\keywords}[1]
{
  \small	
  \textit{Keywords---} #1
}

\newtheorem{proposition}{Proposition}